\documentclass[acmsmall]{acmart}

\AtBeginDocument{%
  \providecommand\BibTeX{{%
    \normalfont B\kern-0.5em{\scshape i\kern-0.25em b}\kern-0.8em\TeX}}}

\usepackage{proof}
\usepackage{wrapfig}
\usepackage{booktabs}   %
\usepackage{subcaption} %
\usepackage{subfiles}
\usepackage{indentfirst}
\usepackage[tableposition=top]{caption}
\usepackage{amsmath}
\usepackage{amsfonts}
\usepackage{mathrsfs}
\usepackage{adjustbox}
\usepackage{mathtools, booktabs}
\usepackage{microtype}
\usepackage{ebproof}
\usepackage{enumitem}
\usepackage{tikz-cd}
\tikzcdset{
  boxedcd/.style={
    every matrix/.append style={
      draw=red,
      thick,
      fill=yellow!50!white,
      rounded corners,
      #1
    },
  },
}
\usepackage{tikz}
\usepackage{multirow}
\usepackage{multicol}
\usepackage{commands}
\usepackage{hyperref}
\usepackage{csvsimple}
\usepackage{hhline}

\usepackage{pgfplotstable}
\usepackage{booktabs}

\usetikzlibrary{arrows}
\usetikzlibrary{positioning}
\usetikzlibrary{tikzmark}

\tikzset{
  treenode/.style = {align=center, inner sep=0pt, text centered, font=\sffamily},
  tnode/.style = {treenode, circle, black, font=\sffamily\tiny, draw=black,
    fill=gray!40, text width=1em},%
  label/.style = {treenode, font=\sffamily, draw=none,
    fill=white, text width=1em},%
  leaf/.style = {treenode, draw=none, font=\sffamily\tiny}%
}

\setcopyright{cc}
\setcctype{by}
\acmDOI{10.1145/3763158}
\acmYear{2025}
\acmJournal{PACMPL}
\acmVolume{9}
\acmNumber{OOPSLA2}
\acmArticle{380}
\acmMonth{10}
\received{2025-03-25}
\received[accepted]{2025-08-12}

\begin{document}

\title{We’ve Got You Covered: Type-Guided Repair of Incomplete Input Generators}

\author{Patrick LaFontaine}
\orcid{0009-0009-4470-3174}
\affiliation{%
  \institution{Purdue University}
  \city{West Lafayette}
  \country{USA}
}
\email{plafonta@purdue.edu}

\author{Zhe Zhou}
\orcid{0000-0003-3900-7501}
\affiliation{%
  \institution{Purdue University}
  \city{West Lafayette}
  \country{USA}
}
\email{zhou956@purdue.edu}

\author{Ashish Mishra}
\orcid{0000-0002-0513-3107}
\affiliation{%
  \institution{IIT Hyderabad}
  \city{Kandi}
  \country{India}
}
\email{mishraashish@cse.iith.ac.in}

\author{Suresh Jagannathan}
\orcid{0000-0001-6871-2424}
\affiliation{%
  \institution{Purdue University}
  \city{West Lafayette}
  \country{USA}
}
\email{suresh@cs.purdue.edu}

\author{Benjamin Delaware}
\orcid{0000-0002-1016-6261}
\affiliation{%
  \institution{Purdue University}
  \city{West Lafayette}
  \country{USA}
}
\email{bendy@purdue.edu}

\begin{abstract}
  Property-based testing (PBT) is a popular technique for
  automatically testing semantic properties of a program, specified as
  a pair of pre- and post-conditions. The efficacy of this approach
  depends on being able to quickly generate inputs that meet the
  precondition, in order to maximize the set of program behaviors that
  are probed. For semantically rich preconditions, purely random
  generation is unlikely to produce many valid inputs; when this
  occurs, users are forced to manually write their own specialized input
  generators. One common problem with handwritten generators is that
  they may be \emph{incomplete}, i.e.,\ they are unable to generate
  some values meeting the target precondition. This paper presents a
  novel program repair technique that patches an incomplete generator
  so that its range includes every valid input. Our approach uses a
  novel enumerative synthesis algorithm that leverages the recently
  developed notion of \emph{coverage types} to characterize the set of
  missing test values as well as the coverage provided by candidate
  repairs. We have implemented a repair tool for OCaml generators,
  called \name{}, and used it to repair a suite of benchmarks drawn
  from the PBT literature.
\end{abstract}



\maketitle

\section{Introduction}
\label{sec:intro}

Property-based testing (PBT) is an increasingly popular methodology
for automatically testing rich semantic properties of systems, with
PBT frameworks targeting most mainstream programming languages,
including Java~\cite{PL+19,RLP+20}, JavaScript~\cite{fastCheck},
Rust~\cite{rustCheck}, Haskell~\cite{quickcheck},
Python~\cite{hypothesis}, Scala~\cite{scalaCheck}, and
OCaml~\cite{QCheck}. In recent years, PBT frameworks have been
effectively applied in a number of real-world settings. Prominent
examples include validating real-world commercial storage
systems~\cite{BJ+21}, ensuring the correctness of formal
specifications against modern architecture and operating system
artifacts~\cite{rems}, and generating executable specifications of
automotive software components~\cite{MAH+17}.

PBT frameworks require two key components from users: \emph{executable
  properties} that capture the expected input-output behaviors of the
system under test (i.e., pre- and post-conditions), and \emph{test
  input generators} that generate random values of the input
types. The values produced by an input generator are used to validate
a system's behaviors, after filtering out any values that do not meet
the stated precondition.
A generator is a nondeterministic program that samples from a space of
values, supplying a \emph{family} of inputs against which programs are
tested. As a simple example, the following generator for integer trees
randomly chooses one of the two constructors of \ocamlinline{int tree}
using a nondeterministic choice operator, $\oplus$, and then
recursively fills in any of its arguments:
\begin{ocaml}
let rec genTree (size : int) : int tree =
  if size <= 0 then Leaf
               else Leaf !$\oplus$! Node(int_gen(), genTree (size - 1), genTree (size - 1))
\end{ocaml}
Many PBT frameworks support automatically deriving a default generator
for an arbitrary algebraic datatype using a similar strategy:
\ocamlinline{genTree} is effectively what is produced by a
\ocamlinline{deriving Arbitrary} clause in QuickCheck, for
instance. Conceptually, a default generator na\"ively samples values
at random: \ocamlinline{genTree n} produces trees of random integers
of height at most \ocamlinline{n}, for example. Unfortunately, many
programs under test impose \emph{sparse preconditions} on their
inputs, i.e., a property that an arbitrary input is unlikely to
satisfy: e.g., valid postal addresses, well-structured XML documents,
red-black trees, or well-typed expressions. If we use
\ocamlinline{genTree} to test a function that expects valid binary
search trees containing at least three elements, for example, we will
have to throw away roughly 95\% of the values it
generates. %
As the precondition grows more restrictive, the overhead of simply
filtering the results of a default generator becomes too great for
most users, especially when testing is part of continuous
integration~\cite{GTHPH+24}. When this occurs, the standard recourse
is to manually write an input generator that produces the desired set
of inputs. This process is unsatisfactory for end-users, however: a
recent study of industrial users of PBT frameworks identified the need
for handwritten generators to ``be a source of friction for many
participants''~\cite{GCDPH+24}, with practitioners stating that
writing generators by hand was a “tedious”
and “high-effort” %
process. %

An important challenge when writing generators tailored to a
particular precondition is identifying which values \emph{not} to
enumerate-- a generator that only produces a restricted set of values
will miss valid parts of the input space, while one that is too
permissive will waste work enumerating terms that are discarded by the
testing framework. While PBT frameworks can report how many generated
terms do not meet a precondition, signaling when a generator is too
permissive, they do not provide similar feedback about the inputs that
an overly restrictive generator will fail to produce. To address this
problem, \citet{poirot} recently proposed \emph{coverage types}, a
type system for reasoning about the values a generator \emph{must}
yield.  Intuitively, a function that fails to type check against a
particular coverage type $\overline{\tau} \rightarrow \nuut{b}{\phi}$
will fail to produce at least one value that satisfies the predicate
$\phi$.
Unfortunately, while coverage types can help developers identify when
the range of a generator is missing certain values, it still falls to
the developer to extend the generator so that its outputs cover those
values.
Simply using the default generator to augment the outputs of an
incomplete generator suffers from the same problems as the na\"ive
sample and filter approach: as our experiments in
\autoref{sec:eval+repair} show, this strategy fails to meaningfully
extend the coverage of an incomplete generator in most
scenarios. Thus, a more targeted approach is needed.

In this paper, we propose an approach that frees the developer from
this obligation by \emph{automatically repairing} an incomplete
generator so that it is complete with respect to a user-specified
property. Our approach uses a novel program synthesis algorithm which
leverages coverage types to build patches that are guaranteed to fill
in any gaps in a generator's coverage. In contrast to the traditional
type-guided program synthesis setting, in which valid solutions are
defined by the \emph{absence} of unwanted/unsafe behaviors, the
success of our repairs is defined by the sorts of behaviors they
\emph{add}. %
This qualitative difference manifests in meaningful ways in the design
of our algorithm: in contrast to the safety specifications used by
traditional deductive synthesis techniques, a top-level specification
of the set of missing values provides limited guidance on how coverage
duties should be distributed among the subexpressions of an incomplete
generator.
On the other hand, it is straightforward to combine partial solutions
that only contribute a piece of the missing coverage to build a
complete solution. Our algorithm leverages this capability to
construct ``minimal'' solutions, i.e., ones that augment the existing
generator with just enough new behaviors to fill in any coverage
gaps-- for almost all the incomplete generators in our experimental
evaluation, 100\% of the values produced by their repaired
counterparts satisfy the target precondition.
As we shall see, our approach can also be used to solve sketch-based
synthesis problems~\cite{Sketch, frangel}, wherein users provide a
generator template comprised of only the control flow structure the
final solution should use, and then rely on our repair algorithm to
generate program fragments that complete this skeleton in a way that
satisfies the target coverage property.

\begin{wrapfigure}{r}{.42\linewidth}
  \centering
  \vspace{-3pt}
  \includegraphics[scale=.2]{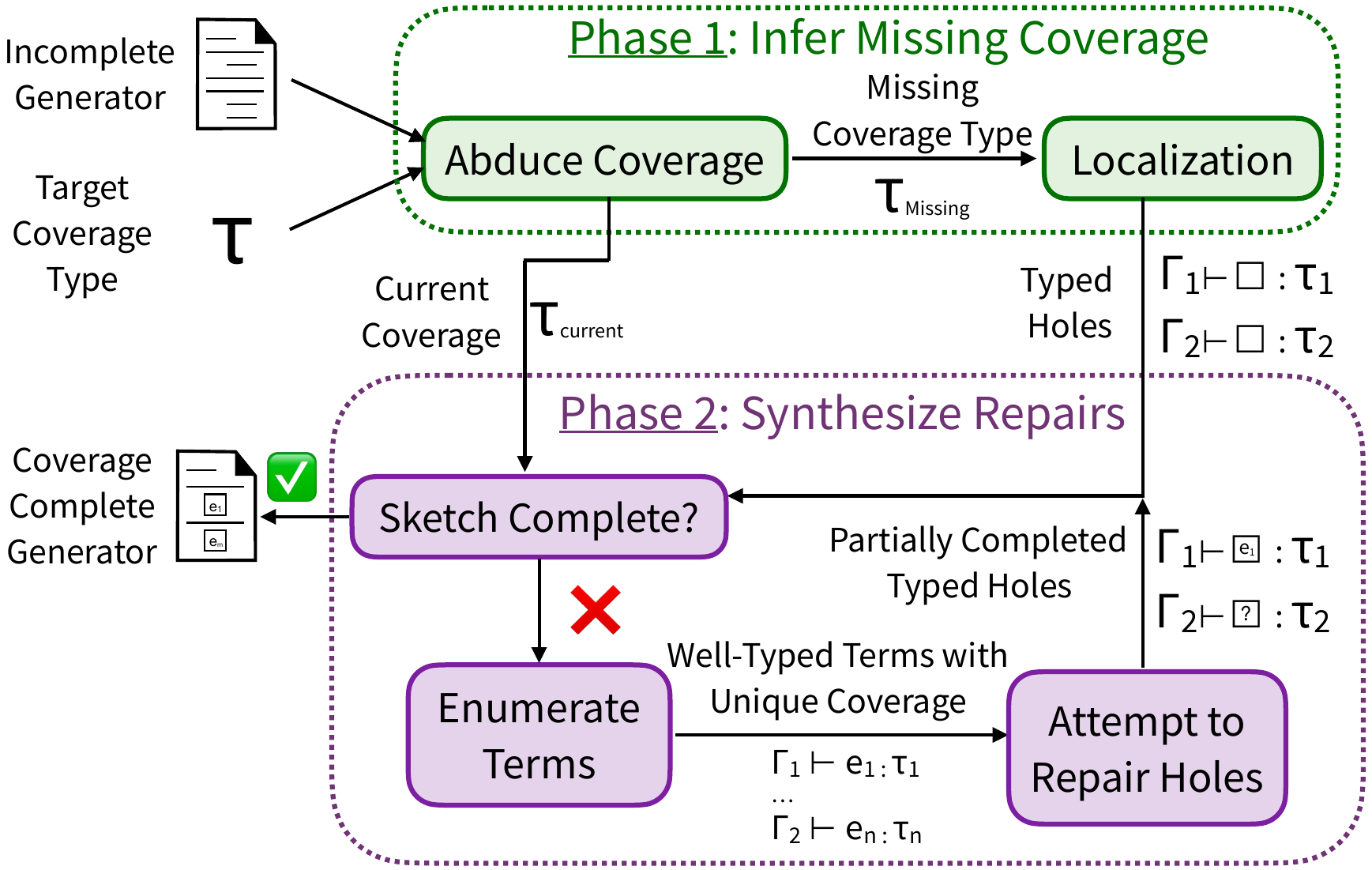}
  \caption{Overview of our proposed pipeline.%
}
\label{fig:pipeline}
\vspace{-2pt}
\end{wrapfigure}

\autoref{fig:pipeline} depicts the high-level workflow of our repair
algorithm and its two main phases. The first
\noindent phase sets up the repair problem, which the second phase
then solves. Our system takes two inputs: an incomplete generator and
a target coverage type specifying the inputs missing from its range.
The first phase begins by characterizing the current and missing
coverage of the input generator using coverage types. It then builds a
program sketch containing typed holes; the typing context of each hole
captures the local variables that can be used to complete that hole.
The algorithm's second phase uses this information to complete
the sketch, employing an enumerative synthesis procedure to find terms
that can be used to patch each of its holes.

In summary, we make the following contributions:
\begin{itemize}

\item Show how coverage types can be used to diagnose and formally
  specify values missing from the range of an incomplete test input
  generator.

\item Present a novel synthesis algorithm that leverages coverage
  types to intelligently repair incomplete test generators.

\item Implement this approach in a tool, \name{}, and demonstrate its
  efficacy by using it to automatically repair a suite of incomplete
  generators for a rich class of datatypes and semantic properties
  drawn from the property-based testing literature.
\end{itemize}

\section{Background and Overview}
\label{sec:overview}

\begin{figure*}[t!]
\begin{tabular}{p{.22\textwidth}|p{.33\textwidth}|p{.35\textwidth}}
  \begin{subfigure}[t]{.13\textwidth}
    \phantomsection{}
    \label{prog:gen+ev+under}
      \begin{ocaml}[fontsize = \footnotesize, linenos = true]
let rec gen0or2 n =
 [0] !$\oplus$! [2]
\end{ocaml}
\end{subfigure}
  &
    \begin{subfigure}[t]{.34\textwidth}
      \phantomsection{}
      \label{prog:gen+ev+over}
      \begin{ocaml}[fontsize = \footnotesize, linenos = true]
let rec genInts n =
 if n == 0 then [int_gen()]
 else
  [int_gen()] !$\oplus$!
  int_gen() :: genInts(n-1)
\end{ocaml}
\end{subfigure}
  &
    \begin{subfigure}[t]{.38\textwidth}
      \phantomsection{}
      \label{prog:gen+ev+close}
      \begin{ocaml}[fontsize = \footnotesize, linenos = true]
let rec genSomeEvens n =
 if n == 0 then [2 * int_gen()]
 else
  (* [2 * int_gen()] !$\oplus$! *)
  2*int_gen() :: genSomeEvens(n-1)
\end{ocaml}
\end{subfigure}
\vspace{0pt}
  \\[3pt] \hline %
  \centering \footnotesize %
  $\ot{\Code{l}}{il}{\Code{l}=[0] \lor \Code{l}=[2]}$
  & \centering \footnotesize %
    <: \quad $\ot{\Code{l}}{il}{0 < \Code{len}(\Code{l}) \leq \Code{n} + 1}$ \quad :>
  & { \centering \footnotesize %
    $\ot{\Code{l}}{il}{\Code{all\_evens}(\Code{l}) \land \Code{len}(\Code{l})
    = \Code{n} + 1}$}
  \\[3pt] \hline %
  \centering \footnotesize %
  $\ut{\Code{l}}{il}{\Code{l}=[0] \lor \Code{l}=[2]}$
  & \centering \footnotesize %
    :> \quad $\ut{\Code{l}}{il}{0 < \Code{len}(\Code{l}) \leq \Code{n} + 1}$ \quad
    <:
  & { \centering \footnotesize %
    $\ut{\Code{l}}{il}{\Code{all\_evens}(\Code{l}) \land \Code{len}(\Code{l}) = \Code{n} +
    1}$}
\end{tabular}
\caption{Three sized generators for non-empty lists of even numbers,
  followed by refinement and coverage types for their bodies. The direction of the subtyping
  relation on the types in each column is included. We use $il$
  as an alias for \ocamlinline{int list}.
}
\label{fig:ev+list+example}
\vspace{-5pt}
\end{figure*}

Before presenting the technical details of our approach, we begin with
a brief review of coverage types and then walk through an end-to-end
example of our repair procedure. \autoref{fig:ev+list+example}
presents three generators for lists of integers: all three are
examples of \textit{sized} generators~\cite{CH+00}, which use a
parameter, in this case \ocamlinline{n}, to bound the number of recursive
calls and thus ensure termination. The first generator in
\autoref{fig:ev+list+example}, \genEvensUnder{}, uses the
nondeterministic choice operator {\small $\oplus$} to randomly produce a
singleton list containing \ocamlinline{0} or \ocamlinline{2}, \genEvensOver{} yields
all non-empty lists of integers whose length is less than \ocamlinline{n+1},
and \genEvensClose{} generates lists of even numbers with
\textit{exactly} \ocamlinline{n+1} elements.

We use these generators to illustrate the difference between coverage
types~\cite{poirot} and traditional refinement
types~\cite{JV21}. Immediately under each generator are a refinement
type, %
{\small $\ot{l}{\Code{int~list}}{\phi}$}, and a coverage type %
{\small $\ut{l}{\Code{int~list}}{\phi}$}. The \emph{qualifiers} $\phi$
of both types constrain the range of the generator above
them. Although the two types are syntactically similar, their semantic
interpretation features an important difference: each refinement type
describes a \emph{superset} of the actual range of the generator above
it, while each coverage type encodes a \emph{subset} of its actual
outputs. This relationship is captured by the subtyping relation for
each kind of type. The refinement type in the middle column is a
supertype of the types on either side of it, so it can also be
assigned to both \genEvensUnder{} and \genEvensClose{}. Coverage types
invert this subtyping relationship: all three coverage types in the
figure can be assigned to \genEvensOver{}, since each type describes a
subset of the values it ``covers''.

Importantly, users will get a type error when checking a generator
against a coverage type whose formula can be satisfied by values that
fall outside its range: while we can type \genEvensUnder{} against %
{\small
  $\ot{l}{\Code{int~list}}{\Code{len}(\Code{l}) = 1 \land 0 \le
    \Code{hd}(\Code{l}) \le 2}$}, we cannot type it against a similar
coverage type, {\small
  $\ut{l}{\Code{int~list}}{\Code{len}(\Code{l}) = 1 \land 0 \le
    \Code{hd}(\Code{l}) \le 2}$}, because \ocamlinline{[1]} is not one
of its outputs. As another example, suppose that a user wants a sized
generator for all non-empty lists that contain even integers, a
property that is captured by the following type signature:%
\vspace{-.4cm}\par\nobreak { \small
\begin{align}
  \Code{n}:\ot{n}{\Code{int}}{\Code{n} \geq 0} \rightarrow%
  \ut{l}{\Code{int~list}}{\lnot \Code{empty}(\Code{l}) \land \Code{len}(\Code{l}) \leq \Code{n} + 1 %
  ~~\land~~ \Code{all\_evens}(\Code{l})}%
  \tag{$\tau_\texttt{Ev}$}\label{eqn:ev+sig+ex}
\end{align}
} %
\noindent The parameter \ocamlinline{n} of this function type has a
standard refinement type that stipulates that the function expects a
non-negative argument. The return type is a coverage type
stipulating that the range of the function includes every non-empty
list of even numbers containing \emph{at most} \ocamlinline{n+1}
elements. Notably, \ref{eqn:ev+sig+ex} is not a valid type for
\genEvensClose{}--- this generator will never return a list with
fewer than \ocamlinline{n+1} elements--- but it is possible to extend this
generator so that it can output such lists by uncommenting the
expression on line 4. We will now describe our approach for
automatically generating these sorts of patches for incomplete
generators.

Our algorithm expects two inputs: a \textit{coverage type} capturing
the set of values the repaired generator must output, e.g.,
\ref{eqn:ev+sig+ex} and an \textit{incomplete generator} to repair.
To illustrate the details of our approach, our walkthrough will use
the following generator, which is an even more incomplete version of
\genEvensClose{} from \autoref{fig:ev+list+example}:\footnote{The
  \Code{err} expression always throws an exception, so \genEvensIncFoot{}
  does not produce any outputs.}%
\phantomsection{}
\label{prog:gen+ev+inc}
\begin{ocaml}
  let rec genEvens!$_\Code{inc}$! (n : int) : int list =
    if n == 0 then err (* base case *) else err (* recursive case *)
  \end{ocaml}

\paragraph{Phase 1: Characterizing Missing Coverage}
Given these inputs, our repair algorithm begins by identifying any
target values not covered by \genEvensInc{}. It does so by inferring a
pair of coverage types, %
{\small $\ut{l}{\Code{int~list}}{\phi_\texttt{cur}}$} and %
{\small $\ut{l}{\Code{int~list}}{\phi_\texttt{need}}$}, that capture
a) the current coverage of the input generator, and b) the coverage
that it is missing, respectively.
Intuitively, b) provides a semantic characterization of the term(s) we
need to synthesize. In the case of \genEvensInc{}, the current
coverage is %
{\small $\ut{l}{\Code{int~list}}{\bot}$}, as the function is not
guaranteed to output \emph{any} values, and the missing coverage type
is simply the return type of \ref{eqn:ev+sig+ex}, i.e., %
\vspace{-.4cm}\par\nobreak %
{ \small
\begin{align}
  \ut{l}{\Code{int~list}}%
  {\lnot \Code{empty}(\Code{l}) \land \Code{len}(\Code{l}) \leq \Code{n} + 1 %
  ~~\land~~ \Code{all\_evens}(\Code{l})}
  \tag{$\tau_\texttt{EvNeed}$}\label{eqn:ev+need+ex}
\end{align}
}
\noindent If we had used \genEvensClose{} instead,
\ref{eqn:ev+need+ex} would be: %
\vspace{-.4cm}\par\nobreak %
{ \small
\begin{align*}
  \small
  \ut{l}{\Code{int~list}}%
  {\Code{n} > 0 \land %
  \Code{len}(\Code{l}) = 1 \land \Code{all\_evens}(\Code{l})}%
\end{align*}
}
\noindent Since the \ocamlinline{else} branch of \genEvensClose{} appends an
even integer to the head of a non-empty list returned by a recursive
call, it always builds a list with at least two elements.

From here, our algorithm builds a \emph{sketch}~\cite{Sketch} of the
complete generator by inserting typed holes into each control flow
path where a patch can be inserted to add coverage. Our motivating
example results in the following sketch with two holes:
\phantomsection{}
\label{prog:gen+ev+sk}
\begin{ocaml}
let rec genEvens!$_\Code{sk}$! (n : int) : int list =
  if n == 0 then !\hole{}$_1 : \text{\ref{eqn:ev+need+ex}}$! (* base case *) else !\hole{}$_2 : \text{\ref{eqn:ev+need+ex}}$! (* recursive case *)
\end{ocaml}
\noindent Our algorithm also attaches a typing context to each hole in
the sketch; intuitively, the typing context of a hole summarizes the
control flow at that program point. The typing contexts for the holes
in our sketch are %
\vspace{-.4cm}\par\nobreak %
{ \small
  \begin{align*}
    \Code{n} : \ot{\;\Code{n}}{\Code{int}}{\Code{n} = 0\;}~\vdash~
    \Code{\hole{}}_1~:~\text{\ref{eqn:ev+need+ex}}
    \quad\quad
    &
      \quad\quad
      \Code{n} : \ot{\;\Code{n}}{\Code{int}}{\Code{n} > 0 \;}~\vdash~
      \Code{\hole{}}_2~:~\text{\ref{eqn:ev+need+ex}}
  \end{align*}
}

\paragraph{Phase 2: Synthesis}
The next phase of our algorithm synthesizes well-typed terms,
$\Gamma_i \vdash e_i~:~\text{\ref{eqn:ev+need+ex}}$ for these holes;
replacing each $\hole{}_i$ in our sketch with $e_i$ will produce a
generator with the target coverage type.  Observe that using the
analogous refinement type as the target type of these terms %
\vspace{-.4cm}\par\nobreak %
{ \small
\begin{align}
  {\ot{l}{\Code{int~list}}%
  {\lnot \Code{empty}(l) \land \Code{len}(\Code{l}) \leq \Code{n} + 1 %
  ~~\land~~ \Code{all\_evens}(\Code{l})}}%
  \tag{$\tau_\texttt{EvBad}$}\label{eqn:ev+need+bad}
\end{align}
}%
\noindent admits numerous solutions that are incongruous with our
intended use of \genEvensSk{} as a test generator. Using
\ref{eqn:ev+need+bad} as the target type for the first hole would
allow $\hole{}_1$ to be filled with any singleton list containing an
even number, including \ocamlinline{[0]}, \ocamlinline{[4]},
\ocamlinline{[n]}, \ocamlinline{[2*n]},
\ocamlinline{[4*int_gen()]}. The first three of these expressions are
consistent with the bias used by many program synthesizers, Occam’s
razor, which prioritizes the ``smallest'' program among candidate
solutions~\cite{Gulwani+11, WDS+17, BPP+20}.

A larger challenge when repairing an incomplete generator is that a
description of the behaviors a patch must add does not provide much
guidance on how to decompose those behaviors into independently
solvable subproblems. To see why, consider how we type check
\genEvensOver{} against the coverage type below it. Neither of the
subexpressions of the {\small $\oplus$} expression in its
\ocamlinline{else} branch check against this type, because neither
individually covers all the values it stipulates --- indeed, if either
did so, the other expression would be redundant! In general, when
typing an expression of the form %
{\small $\Code{e_1} \oplus \Code{e_2}$} against a coverage type %
{\small $\ut{l}{\Code{b}}{\phi}$}, we cannot simply independently
check \ocamlinline{e}$_\Code{1}$ and \ocamlinline{e}$_\Code{2}$
against that type. Instead, we need to come up with types %
{\small $\ut{l}{\Code{b}}{\phi_1}$} and %
{\small $\ut{l}{\Code{b}}{\phi_2}$ } to check
\ocamlinline{e}$_\Code{1}$ and \ocamlinline{e}$_\Code{2}$ against, and
then check that the combined coverage of those types is sufficient,
i.e.  %
{\small
  $\ut{l}{\Code{b}}{\phi_1 \lor \phi_2} <:
  \ut{l}{\Code{b}}{\phi}$}. %
When type checking %
{\small $\Code{e_1} \oplus \Code{e_2}$}, we can use
\ocamlinline{e}$_\Code{1}$ and \ocamlinline{e}$_\Code{2}$ to help
infer %
{\small $\ut{l}{\Code{b}}{\phi_1}$} and %
{\small $\ut{l}{\Code{b}}{\phi_2}$}, but a top-down, type-directed
synthesis algorithm does not have either \ocamlinline{e}$_\Code{1}$
or \ocamlinline{e}$_\Code{2}$ in hand; it is responsible for
generating both terms from a type. Unfortunately, there are many
possible ways to partition the coverage responsibilities of a {\small
  $\oplus$} expression between its subexpressions, each of which
results in a different set of synthesis goals, and it is not obvious
how to choose between these partitionings.

\begin{wrapfigure}{r}{.44\linewidth}
  \vspace{-.275cm}
  \small
  \addtolength{\tabcolsep}{-4pt}
  \begin{tabular}{rl}
    $\Sigma_0 \equiv \{ $
    & \ocamlinline{[0]}, \ocamlinline{[n]}, \ocamlinline{1}, \ocamlinline{3}, \ocamlinline{[ ]}, \ocamlinline{Leaf}, \\
    & \ocamlinline{[int_gen()]}, $\; \ldots \; \}$ %
    \\[3pt] \hline \rule{0pt}{1.015\normalbaselineskip} %
    $\Sigma_1 \equiv \{  $
    & \ocamlinline{2*int_gen()}, \ocamlinline{genEvens(n-1)}, $\; \ldots\; \}$  %
    \\[3pt] \hline \rule{0pt}{1.015\normalbaselineskip} %
    $\Sigma_2 \equiv \{ $
    & \ocamlinline{0 :: genEvens(n-1)}, \\
    & \ocamlinline{n :: genEvens(n-1)}, \\
    & \ocamlinline{int_gen() :: genEvens(n-1)}, $\; \ldots \; \}$  %
    \\[3pt] \hline \rule{0pt}{1.015\normalbaselineskip} %
    $\Sigma_3 \equiv \{  $
    &\ocamlinline{genEvens(n-1) ++ genEvens(n-1)}, \\
    & \ocamlinline{2*int_gen() :: genEvens(n-1)}, $\; \ldots \; \}$
  \end{tabular}
  \vspace{-.225cm}
  \caption{Example sets of enumerated terms;
    elements of $\Sigma_i$ cost less than elements of
    $\Sigma_{i+1}$. %
  }
  \label{fig:enum+terms}
  \vspace{-.45cm}
\end{wrapfigure}

As a consequence, our synthesis procedure instead adopts a bottom-up
approach: iteratively generating a set of partial solutions that can
be combined to construct a complete answer. Our algorithm maintains a
pool of candidate terms that it uses to generate new terms; this pool
grows as the algorithm proceeds. At each iteration of the loop, the
algorithm uses a \emph{syntactic} cost function to prioritize the
generation of certain terms. \autoref{sec:synthesis} provides more
detail on our cost function, but intuitively, smaller terms and terms
with more coverage, like generators and recursive calls, have lower
cost. \autoref{fig:enum+terms} provides some examples of terms at
different cost levels for our running example.
After generating all the terms at the current cost threshold, our
algorithm infers a coverage type for each expression, and uses this
\emph{semantic} information to prune out any terms that are unsafe,
not useful, or redundant. In the case of terms containing a recursive
call, for example, type checking ensures that the first argument to
each recursive call is structurally decreasing, ensuring that a
generator using such a term will terminate. Our algorithm also uses
the inferred types to safely discard any terms that do not provide new
coverage. If the pool of candidates already contains the term
\ocamlinline{int_gen()}, for example, there is no reason to add
\ocamlinline{int_gen()+1} or \ocamlinline{int_gen()+int_gen()} to it:
all of these expressions generate the same terms, and thus have the
exact same coverage type. We only add terms that satisfy these sorts
of semantic conditions to the pool of enumerated terms.

\begin{figure}[!t]
  \centering
  \begin{tikzpicture}[->,>=stealth',level 1/.style={sibling distance = .5\linewidth},
    level 2/.style={sibling distance = .25\linewidth}]
    \node  { \footnotesize %
      $\Gamma_2\;\vdash$ \ocamlinline[fontsize = \footnotesize]{genList n} $~:~ \ut{l}{\Code{int~list}}{\top}$
    }
    child{ node[yshift=3mm]  {
        \begin{minipage}{.45\linewidth}
        \footnotesize
        $\Gamma_2\;\vdash$ \ocamlinline[fontsize = \footnotesize]{[2*int_gen()]} $~:~$ \\
        $\ut{l}{\Code{int~list}}%
        {\Code{len}(\Code{l}) = 1 \land \Code{even(hd}(\Code{l}))}$%
        \end{minipage}
      }
      child{ node[yshift=1mm]  {
          \begin{minipage}{.4\linewidth}
            \footnotesize
          $\Gamma_2\;\vdash$ \ocamlinline[fontsize = \footnotesize]{[0]} $~:~$\\
          $\ut{l}{\Code{int~list}}%
          {\Code{len}(\Code{l}) = 1 \land \Code{hd}(\Code{l}) = 0}$%
          \end{minipage}
        }
        edge from parent [->] node [above, rotate=35] {$\sqsupseteq$}
      }
      child{ node[yshift=-8mm, xshift=-10mm]  {
          \begin{minipage}{.4\linewidth}
          \footnotesize
          $\Gamma_2\;\vdash$ \ocamlinline[fontsize = \footnotesize]{[2*n]} $~:~$ \\
          $\ut{l}{\Code{int~list}}%
          {\Code{len}(\Code{l}) = 1 \land \Code{hd}(\Code{l}) = 2*n}$%
        \end{minipage}
      }
      edge from parent [->] node [above, rotate=-70] {$\sqsubseteq$}
      }
      edge from parent [->] node [above, rotate=30] {$\sqsupseteq$}
    }
    child{ node[yshift=3mm, xshift=-10mm]  {
        \begin{minipage}{.45\linewidth}
        \footnotesize
        $\Gamma_2\;\vdash$ \ocamlinline[fontsize = \footnotesize]{int_gen() :: genEvens(n-1)} $~:~$ \\
        $\ut{l}{\Code{int~list}}%
        {\Code{len}(\Code{l}) \le \Code{n} + 1\land \phi(\Code{l})}$%
      \end{minipage}
      }
      child{ node[yshift=2mm]  {
          \begin{minipage}{.4\linewidth}
            \footnotesize
            $\Gamma_2\;\vdash$ \ocamlinline[fontsize = \footnotesize]{2*int_gen() :: genEvens(n-1)} $~:~$ \\
            $\ut{l}{\Code{int~list}} %
            {\Code{len}(\Code{l}) \le \Code{n} + 1 \land \Code{even}(\Code{hd}(\Code{l})) \land \phi(\Code{l})}$%
          \end{minipage}
        }
        child{ node[yshift=2mm]  {
          \begin{minipage}{.4\linewidth}
            \footnotesize
            $\Gamma_2\;\vdash$ \ocamlinline[fontsize = \footnotesize]{0 :: genEvens(n-1)} $~:~$ \\
            $\ut{l}{\Code{int~list}}%
            {\Code{len}(\Code{l}) \le \Code{n} + 1 \land
              \Code{hd}(\Code{l}) = 0 \land \phi(\Code{l}) }$%
          \end{minipage}
        }
        edge from parent [->] node [above, rotate=90] {$\sqsupseteq$}
        }
        edge from parent [->] node [above, rotate=90] {$\sqsupseteq$}
      }
      edge from parent [->] node [above, rotate=-30] {$\sqsubseteq$}
    }
  ;
\end{tikzpicture}
\caption{A subset of the join semi-lattice built for $\hole{}_2$ in
  \genEvensSk{}, where
  $\Gamma_2 \equiv \Code{n}:\ot{\;\Code{n}}{\Code{int}}{\Code{n} >
    0\;}$ and $\phi(\Code{l}) \equiv %
  \lnot \Code{empty}(\Code{tail(l)}) \land %
  \Code{len}(\Code{tail(l)}) \leq (\Code{n}-1) + 1 \land %
  \Code{all\_evens}(\Code{tail(l)})$.}
\label{fig:even+lattice}
\end{figure}

The final step in our algorithm's enumeration loop checks if a valid
completion for any of the holes in the sketch has been found. To do
so, it maintains a set of enumerated terms with the same base type as
the hole, under the typing context for that hole. This set is
partially ordered by the subtyping relation on the types inferred for
elements by our type inference algorithm,
\typeInfer{}:%
\begin{align*}
  e_1 \sqsubseteq e_2 \equiv
  & \typeInfer{}(\Gamma_k,\; e_1) <: \typeInfer{}(\Gamma_k,\; e_2)
\end{align*}
\noindent \autoref{fig:even+lattice} shows an example of part of this
poset for $\hole{}_2$ in \genEvensSk{}. As we have
seen, if the terms \ocamlinline{e}$_\Code{1}$ and \ocamlinline{e}$_\Code{2}$ have
the coverage types $\tau_1$ and $\tau_2$ where $\tau_1 <: \tau_2$,
then \ocamlinline{e}$_\Code{2}$ is only guaranteed to generate a subset of
the outputs of \ocamlinline{e}$_\Code{1}$. Thus, this poset tracks the
relative coverages of the candidate solutions (as determined by
\typeInfer{}) our algorithm has enumerated so far.
The top element in this poset is the default generator for our target
type, capable of enumerating every list of integers. Its two children
only produce a subset of its outputs; they are sibling nodes because
neither is a subtype of the other, i.e., neither's outputs subsumes
the other's. Importantly, this poset forms a join-semilattice: given
any two terms, we can build a term that covers both sets of inputs by
joining them together via our nondeterministic choice operator: %
{\small
  $\Code{e_1} \sqsubseteq (\Code{e_1} \oplus \Code{e_2}) \sqsupseteq
  \Code{e_2}$}. Our implementation of this poset does not need to
maintain these sorts of elements, as it can always use {\small
  $\oplus$} to reconstruct them on demand: as a consequence, there is
no need for \autoref{fig:even+lattice} to explicitly include %
\vspace{-.4cm}\par\nobreak %
{ \small
\begin{align*}
  \Code{n}:\ot{\;\Code{n}}{\Code{int}}{\Code{n} > 0\;}  \vdash %
  \text{\ocamlinline{[0]}} \oplus\text{\ocamlinline{[2*n]}}~:~\ut{l}{\Code{int~list}}%
  {\Code{len(l)} = 1 \land \Code{hd(l)} = 0 %
  \lor \Code{len(l)} = 1 \land \Code{hd(l)} = \text{\Code{2*n}}}
\end{align*}
}

To check if it has found a solution for a hole, our algorithm first
walks down this lattice looking for an element with the same type as
the hole, returning that element as the solution if so. The poset
corresponding to \autoref{fig:even+lattice} for $\hole{}_1$ contains a
such direct solution, for example: %
\vspace{-.4cm}\par\nobreak %
{\small
  \begin{align}
    \Code{n}~:~\ot{\;\Code{n}}{\Code{int}}{\Code{n} =
    0\;}~\vdash~\text{\ocamlinline{[2*int_gen()]}}~:~\text{\ref{eqn:ev+need+ex}}
    \tag{\Code{p$_1$}}
    \label{eqn:patch+1}
  \end{align}
} %
If a direct solution is not available, our algorithm attempts to build
a solution by joining together all the elements that would be
immediate subchildren of a hypothetical expression with the target
coverage type. %
While there is no immediate solution to $\hole{}_2$ in
\autoref{fig:even+lattice}, it does contain two expressions that would
be direct children of such a node: %
\ocamlinline{[2*int_gen()]} and %
\ocamlinline{2*int_gen() :: genEvens(n-1)}. The join of these
expressions provides precisely the coverage required by $\hole{}_2$: %
\vspace{-.4cm}\par\nobreak %
{\small
  \begin{align}
    \Code{n}~:~\ot{\;\Code{n}}{\Code{int}}{\Code{n} > 0\;}~\vdash~
    \text{\ocamlinline{[2*int_gen()]}} ~~\oplus~~ \text{\ocamlinline{2*int_gen() :: genEvens(n-1)}}
    ~:~\text{\ref{eqn:ev+need+ex}}
    \tag{\Code{p$_2$}}
    \label{eqn:patch+2}
  \end{align}
} %
\noindent Replacing the holes in \genEvensSk{} with
\ref{eqn:patch+1} and \ref{eqn:patch+2} results in a complete
generator that is identical to a version of \genEvensClose{} with its
fourth line uncommented.

We pause here to highlight the distinguishing features of our
algorithm: the first is its use of the coverage type
\ref{eqn:ev+need+ex} to precisely characterize the behaviors a repair
needs to add in order to make \Code{genEvens$_\Code{inc}$}
complete. While \ref{eqn:ev+need+ex} provides a semantic specification
for the top-level synthesis problem, it does not provide much guidance
on how to decompose that problem into independently solvable subgoals,
e.g., when patching $\hole{}_2$. Thankfully, the nondeterministic
nature of input generators enables the second key feature of our
bottom-up synthesis algorithm, its ability to use $\oplus$ to combine
partial patches into a complete solution, a capability that it used to
generate \ref{eqn:patch+2}.

\section{Language}
\label{sec:language}

\begin{figure}[t!]
{\small
    \begin{alignat*}{2}
    \text{\textbf{Variables }}& \quad &\quad& x, f, u, ... \\[-2pt]
    \text{\textbf{Data constructors }}& \quad &d ::= \quad & () ~|~ \S{true} ~|~ \S{false} ~|~ \S{O} ~|~ \S{S} ~|~ \S{Cons} ~|~ \S{Nil} ~|~ \S{Leaf} ~|~ \S{Node} \\[-2pt]
    \text{\textbf{Constants }}& \quad &c ::= \quad & \mathbb{B} ~|~ \mathbb{N} ~|~ \mathbb{Z} ~|~ \ldots ~|~ d\ \overline{c}\\[-2pt]
    \text{\textbf{Operators }}& \quad & op ::= \quad &d ~|~ {+} ~|~ {==} ~|~ {<} ~|~ \Code{mod} ~|~ \randomnat ~|~  \Code{int\_gen} ~|~ ...\\[-2pt]
    \text{\textbf{Values }}& \quad  & v ::= \quad & c ~|~ op ~|~ x ~|~ \zlam{x}{t}{e} ~|~ \zfix{f}{t}{x}{t}{e}  \\[-2pt]
    \text{\textbf{Terms} and} &
    \quad & e, \textcolor{DeepGreen}{s} ::=\quad & v ~|~ \S{err}
    ~|~ \zlet{x}{e}{e}
    ~|~ \zlet{x}{op\ \overline{v}}{e}
    ~|~ \zlet{x}{v\ v}{e} \\[-2pt]
    & & & ~|~ \match{v} \overline{d\ \overline{y} \to e} \\[-2pt]
    \text{\textcolor{DeepGreen}{\textbf{Incomplete Terms}}}
    &\quad&\quad& ~|~ \textcolor{DeepGreen}{\hole{} ~:~ \nuut{\textit{b}}{\phi}}   \\[-2pt]
    \text{\textbf{Base Types}}& \quad & b  ::= \quad &  unit ~|~ bool ~|~ nat ~|~ int ~|~  b\ list ~|~ b\ tree ~|~ \ldots \\[-2pt]
    \text{\textbf{Basic Types}}& \quad & t  ::= \quad & b ~|~ t\, {\shortrightarrow}\,t \\[-2pt]
    \text{\textbf{Method Predicates }}& \quad & mp ::= \quad & \I{emp} ~|~ \I{hd} ~|~ \I{mem} ~|~ ...\\[-2pt]
    \text{\textbf{Literals }}& \quad & l ::= \quad & c ~|~ x \\[-2pt]
    \text{\textbf{Propositions}}& \quad & \phi ::= \quad &  l ~|~ \bot ~|~ \top_b  ~|~ op(\overline{l}) ~|~ mp(\overline{x}) ~|~ \neg \phi ~|~ \phi \land \phi ~|~ \phi \lor \phi ~|~ \phi \impl \phi ~|~  \forall u{:}b.\; \phi ~|~ \exists u{:}b.\; \phi \\[-2pt]
    \text{\textbf{Refined Types}}& \quad &\tau  ::= \quad &  \nuut{\textit{b}}{\phi} ~|~ \nuot{\textit{b}}{\phi} ~|~  x{:}\tau{\shortrightarrow}\tau
    \\[-2pt]
    \text{\textbf{Type Contexts}}& \quad &\Gamma ::= \quad & \emptyset ~|~ \Gamma, x{:}\tau
  \end{alignat*}
}
\vspace{-.8cm}
\caption{\langname{} syntax.}\label{fig:syntax}
\end{figure}

To formalize our type-based approach to test generator synthesis and
repair, we use \langname{}, a slightly modified version of
\origlangname{}, a core calculus for input generators introduced by
\citet{poirot}. This section reviews the key features of that original
calculus, highlighting our extensions along the way. The syntax of
\langname{} is shown in \autoref{fig:syntax}. The language is a
call-by-value lambda-calculus with pattern-matching, inductive
datatypes, and recursive functions. Programs are written in monadic
normal-form (MNF)~\cite{JO94}, a variant of A-Normal Form
(ANF)~\cite{FSDF93} that allows nested let-bindings. \langname{} is
equipped with generators for numeric types-- \sCode{nat\_gen} and
\sCode{int\_gen}-- which can evaluate to any number in their range with
nonzero probability. These built-in generators suffice to express
additional nondeterministic behaviors: the $\oplus$ choice operator,
for example, can be defined as: %
\vspace{-.4cm}\par\nobreak %
{ \small
\begin{align*}
    e_1\oplus e_2 \equiv &\ \zlet{\Code{n}}{\randomnat\; ()\
                           \Code{mod}\ 2}{\match{\Code{n}} 0 \to e_1 ~|~\_ \to e_2}
\end{align*}
}
\noindent Like its predecessor, \langname{} does not include operators
to bias how often values are produced, e.g., QuickCheck's
\sCode{frequency}; including such an operator would not fundamentally
impact the guarantees we provide for synthesized
generators. \langname{} is equipped with a completely standard
small-step operational semantics, $e \hookrightarrow e'$, that mirrors that
of \origlangname{}.

\begin{wrapfigure}{r}{.3\linewidth}
  \vspace{-.25cm}
  $\begin{prooftree}
    \hypo{v \models \phi}
    \infer1[\textcolor{DeepGreen}{\textsc{EHole}}]
    {
      \hole{}~:~\nuut{b}{\phi} \hookrightarrow v}
  \end{prooftree}$ %
  \vspace{-.30cm}
\end{wrapfigure}
The only addition \langname{} makes to \origlangname{} is an
additional syntactic category of \emph{incomplete terms}
$\textcolor{DeepGreen}{s}$; these terms may contain one or more typed
\emph{holes}, \hole{}$~:~\nuut{b}{\phi}$.  Semantically, holes can
evaluate to any value satisfying $\phi$ (\textsc{EHole}), and thus act
as a kind of semantic placeholder for a complete patch.
Syntactically, our algorithm uses holes to identify program points at
which repairs can be inserted. Given an incomplete program $s$ with
$j$ holes, we write $s[\overline{e}]$ to denote the complete program
where the $i^\text{th}$ hole has been replaced by $e_i$. The output of
our repair algorithm is a \emph{syntactically} complete
\origlangname{} program, i.e., it does not contain any holes, that is
also \emph{semantically} complete, meaning it can produce all inputs
satisfying the target property. %

\subsection{Type System}
\label{sec:type+system}

\langname{} inherits the type system of its predecessor; like
\origlangname{}, \langname{} has three categories of types: \emph{base
  types}, \emph{basic types}, and \emph{refined types}.  Base types
($b$) include primitive types, e.g., \sCode{unit} and \sCode{bool},
and inductive datatypes, e.g., \sCode{int list} and \sCode{bool
  tree}. Basic types ($t$) extend base types with function types. As
in other refinement type systems, refined types ($\S{\tau}$) qualify
base types with predicates in a decidable fragment of first-order
logic (FOL). In \langname{}, however, type refinements have two
distinct modalities: as we saw in \autoref{sec:overview}, the
qualifiers of coverage types ($\nuut{b}{\phi}$) identify a
\emph{subset} of the values a nondeterministic expression must be able
to evaluate to, while the qualifiers of refinement types
($\nuot{b}{\phi}$) characterize a \emph{superset} of the values an
expression may evaluate to. In order to express rich shape properties
over inductive datatypes, we allow propositions to reference
\emph{method predicates}, boolean-valued functions on inductive
datatypes like $\I{emp}$, $\I{hd}$, and $\I{mem}$. Using such
predicates, it is straightforward to generate verification conditions
that can be handled by an off-the-shelf theorem prover like
Z3~\cite{de2008z3}. In order to ensure that type checking is
decidable, our type system restricts refinements to effectively
propositional (EPR) sentences (i.e., prenex-quantified formulae of the
form $\exists^*\forall^*\varphi$ where $\varphi$ is quantifier-free).
Following \origlangname{}, our type system allows function parameters
to be qualified by refinements that specify when it is safe to apply a
test generator, while a generator's return type is qualified using a
coverage type that characterizes the values it is guaranteed to
produce.

\begin{figure}[t!]
  {\small
{\normalsize
\begin{flalign*}
 &\text{\textbf{Typing }} & \fbox{$\Gamma \covervdash s : \tau$}
\end{flalign*}
}
\\ \
  \begin{prooftree}
    \hypo{\Gamma \vdashunder \nuut{b}{\phi}}
    \infer1[\textcolor{DeepGreen}{\textsc{THole}}]{ \Gamma \covervdash
      \hole{}~:~\nuut{b}{\phi} : \nuut{b}{\phi} }
  \end{prooftree}
  \quad
  \begin{prooftree}
    \hypo{
      \Gamma \covervdash v : \tau_v \hspace{0.85em}
      \overline{\Gamma, \overline{{y}{:}\tau_y} \covervdash  d_i(\overline{{y}}) : \tau_v} \hspace{0.85em}
      \overline{\Gamma, \overline{{y}{:}\tau_y} \covervdash e_i : \tau_i} \hspace{0.85em}
      \overline{\Gamma \vdashunder \tau_i}
      }
    \infer1[\textsc{TMatch}]{
      \Gamma \covervdash \match{v} \overline{d_i\ \overline{y} \to e_i} ~~:~~ \bigvee_i \tau_i}
  \end{prooftree}
\\ \ \\ \ \\
  \begin{prooftree}
\hypo{
\Gamma,\; x{:}\nuot{b}{\phi},\; f{:}\nuotxarr{x}{b}{\nu \prec x~\land~ \phi}\tau ~~\covervdash~~ e : \tau \quad
\Gamma \vdashunder \nuotxarr{x}{b}{\phi}\tau
}
\infer1[\textsc{TFix}]{
  \Gamma \vdash \zfix{f}{\tau}{x}{b}{e} ~~:~~ \nuotxarr{x}{b}{\phi}\tau
}
\end{prooftree}
  }
  \vspace{-.15cm}
\caption{Selected \langname{} typing rules.}
\label{fig:typing+rule}
\vspace{-.5cm}
\end{figure}

\autoref{fig:typing+rule} presents selected typing rules for
\langname{}. The newly added typing rule for holes, \textsc{THole},
reflects the intuition that a hole is an oracle that can produce any
value satisfying the qualifier of its annotated type. The sole premise
of \textsc{THole} is a $\vdashunder$ judgment that ensures that a hole
is annotated with a well-formed type, e.g., that $\phi$ does not
include any free variables.\footnote{The appendices of the full version
  of the paper~\cite{cobb:full} include the complete set of typing
  rules, auxiliary judgements, and a proof of type soundness.}
The typing rule for \sCode{match} expressions, \textsc{TMatch},
reflects the fact that the coverage provided by pattern matching is the union
($\lor$) of the coverages of its branches, each of which may
contribute a different set of values. This is in contrast to how
branching control flow structures are treated in standard refinement
type systems, where each branch can be independently checked against
the type of the overall expression.
The type system of \langname{} enforces the same high-level properties
as that of \origlangname{}: the typing rule for recursive functions,
\textsc{TFix}, for example, uses a well-founded relation on the first
argument of a function to ensure that it terminates.
The remaining typing rules are identical to those of \origlangname{}
and are similar to other refinement type systems~\cite{JV21}.

For the purposes of automatic test generator repair, the key property
enforced by this type system is that a well-typed term \sCode{e} with
the type $\nuut{b}{\phi}$ can evaluate to every value satisfying $\phi$:
\begin{theorem}[Type Soundness~\cite{poirot}]
  \label{cor:typed+complete}
  A well-typed test generator of type
  $ \covervdash f \;:\; \overline{x_i: \nuot{b_i}{\phi_i}}
  \shortrightarrow \nuut{b}{\phi} $, when applied to well-typed
  arguments $\overline{\covervdash v_i~:~\nuot{b_i}{\phi_i}}$, can
  evaluate to every value satisfying
  $\phi[\overline{x_i\mapsto v_i}] $:
  $\forall v.\; \phi[\overline{x_i\mapsto v_i}, \nu\mapsto v] \impl
  f\; \overline{v_i} \hookrightarrow^* v$%
\end{theorem}
\langname{} is also equipped with a decidable bidirectional typing
algorithm whose type synthesis (\typeInfer{}) and type checking
subroutines will play key roles in the repair algorithm we now
present.

\section{Input Generator Repair}
\label{sec:algorithm}

\begin{algorithm}[t!]
  \small
  \Params{$\; s$: incomplete program, %
    $\Gamma$: typing context for $s$, %
    $\nuut{b}{\psi}$: target coverage type for $s$} %
  \Output{\phantom{a} Coverage complete repaired program $e$ such that
    $\Gamma \covervdash e \; : \; \nuut{b}{\psi}$}
  $\nuut{b}{\psi_\texttt{cur}} \leftarrow \typeInfer{}(\Gamma, s)$; %
  \CodeComment{Infer initial coverage of $s$}

  $\nuut{b}{\psi_\texttt{need}} \leftarrow \generalize(\Gamma,
  \nuut{b}{\psi_\texttt{cur}}, \nuut{b}{\psi})$; %
  \CodeComment{Abduce missing coverage}

  $(s', \overline{\Gamma_j\vdash\hole{}_j~:~\nuut{b}{\psi_j}}) \leftarrow
  \localize(\Gamma, s, \nuut{b}{\psi_\texttt{need}})$; %
  \CodeComment{Identify repair locations}

  \Return{$\synthesize(\Gamma, s', %
    \overline{\Gamma_j\vdash\hole{}_j~:~\nuut{b}{\psi_j}}, %
    \nuut{b}{\psi_\texttt{cur} \lor \psi_\texttt{need}})$}; \CodeComment{Synthesize patches for holes}
  \caption{The high-level coverage repair algorithm (\complete{})}
  \label{algo:Complete}
\end{algorithm}

Our top-level repair algorithm, shown in \autoref{algo:Complete},
closely follows the workflow depicted in \autoref{fig:pipeline}. Most
of its functionality is delegated to three key subroutines
(\generalize{}, \localize{}, and \synthesize{}); this section presents
the important details of these subroutines, focusing in particular on
\synthesize{}. \complete{} takes the body of the target generator $s$
(potentially with user-provided holes), a typing context $\Gamma$, and
the target coverage type $\nuut{b}{\psi}$. The algorithm is
additionally parameterized over several ingredients that it uses to
construct repairs: a collection of typed components that \synthesize{}
uses to enumerate terms, a syntactic cost function used to prioritize
which terms to enumerate, an upper bound on the cost of enumerated
patches, the set of method predicates used in the types of those
components and by \generalize{} to characterize missing coverage, and
axioms characterizing the semantics of those method predicates. To
avoid cluttering our discussion, we leave these parameters implicit in
the definition of \complete{} and its subroutines.

\complete{} begins by inferring two coverage types,
$\nuut{b}{\psi_\texttt{cur}}$ (line $1$) and
$\nuut{b}{\psi_\texttt{need}}$ (line $2$). The former characterizes
the current coverage of $s$, and the latter describes the coverage
that $s$ lacks. Next, \complete{} uses \localize{} (line $3$) to
construct a sketch $s'$ that contains holes at each location in $s$
where coverage should be added, as well as a context and type for each
hole, $\overline{\Gamma_j\vdash
  \hole{}_j : \nuut{b}{\psi_j}}$. \complete{} then constructs the
final generator by using \synthesize{} to patch each hole in $s'$
(line $4$).

\subsection{Inferring Missing Coverage}
\label{sec:Generalization}

\begin{figure}[t]
  \begin{subfigure}[t]{.31\linewidth}
    \vskip 0pt
    \begin{ocaml}[fontsize=\footnotesize]
if n == 0
 then [ ]
 else
  let h = int_gen() in
  let t = genIntList(n-1) in
   h :: t
 \end{ocaml}
 \vspace{-3pt}
\caption{\ocamlinline{genIntList}}
\label{fig:sized+list+impl}
\end{subfigure}
\vspace{-2pt}
\begin{subfigure}[t]{.67\linewidth}
  \vskip 0pt
  \footnotesize
  $\psi \equiv \Code{len(l)} \le \Code{n}$ \\[-2pt]
  \noindent\rule{\textwidth}{1pt}
  $
  \psi_\texttt{cur} \equiv
  \begin{aligned}
    &\Code{n}=0 \implies\Code{empty(l)} \\[-2pt]
    \land~ & \Code{n}\neq 0 \implies \exists h. \exists t. \Code{len}(t) \le \Code{n-1} \land
    \Code{hd(l)}=h \land \Code{tl(l)}=t
  \end{aligned}$ \\[3pt]
  \noindent\rule{\textwidth}{1pt}
  $\psi_\texttt{need} \equiv \Code{len(l)} = 0 \land \Code{len(l)} \le
  \Code{n} $
  \caption{Qualifiers of the target, inferred and abduced coverage of
    \ocamlinline{genIntList}.}
\label{fig:sized+list+types}
\end{subfigure}
\vspace{-2pt}
\caption{An incomplete generator for length-bounded lists of integers
  and coverage type qualifiers capturing its target %
  {\footnotesize $\ut{l}{\Code{int~list}}{\psi}$}, current %
  {\footnotesize $\ut{l}{\Code{int~list}}{\psi_\Code{cur}}$} and
  missing %
  {\footnotesize $\ut{l}{\Code{int~list}}{\psi_\Code{need}}$}
  coverage. }
\label{fig:sized+list+ex}
\vspace{-2pt}
\end{figure}

The \generalize{} subroutine infers a coverage type that captures a
set of values missing from the range of a generator. Notably,
\generalize{} may return a coverage type that is more general than
necessary, i.e., it may represent a superset of the values needed to
complete a generator. To motivate why, consider the incomplete
generator for length-bounded lists of integers shown in
\autoref{fig:sized+list+impl}, \ocamlinline{genIntList}. To the right
of \ocamlinline{genIntList} are qualifiers for the coverage types that
\complete{} will use to characterize the target ($\psi$), current
($\psi_\texttt{cur}$), and missing ($\psi_\texttt{need}$) coverage of
\ocamlinline{genIntList}. This generator always returns a list with
\emph{exactly} \ocamlinline{n} members, so it will fail to type check
against a coverage type qualified with $\psi$, which stipulates that
\ocamlinline{genIntList} should return every list containing \emph{at
  most} \ocamlinline{n} integers.  To perform this check, our type
checker infers a type for \ocamlinline{genIntList} using\typeInfer{},
which produces a coverage type with the qualifier
$\psi_\texttt{cur}$. Note that \typeInfer{} always infers the most
precise type it can, so the complexity of $\psi_\texttt{cur}$ is
commensurate with the definition of \ocamlinline{genIntList}, e.g.,
the number of its control flow paths and the components it uses. Thus,
while we could capture the missing coverage of
\ocamlinline{genIntList} by taking the intersection of $\psi$ and
$\psi_\texttt{cur}$, i.e., $\psi \land \lnot \psi_\texttt{cur}$, the
resulting type is overly complex and does not account for the coverage
of the components used by \synthesize{} to construct repairs.
Instead, \complete{} uses \generalize{} to find a simpler, but still
precise characterization of the missing coverage that also aligns with
the space of possible patches explored by \synthesize{}. In the case
of \ocamlinline{genIntList}, for example, \generalize{} identifies a
qualifier $\psi_\texttt{need}$ which succinctly captures the coverage
that needs to be added to \ocamlinline{genIntList}.

\generalize{} is parameterized over a finite set of atomic formulas,
and explores candidate solutions of the form
$\bigvee (\bigwedge\overline{\phi} \land \bigwedge \overline{\lnot
  \phi}) \land \psi$, where $\phi$ are drawn from this set.  If this
set contains %
{\small $\Code{len(l)} = 1$}, {\small $\Code{empty(l)}$}, %
{\small $\Code{all\_evens(l)}$}, and {\small $\Code{n} = 0$}, the set
of qualifiers considered by \generalize{} for \ocamlinline{genEvens}
includes:
\begin{itemize}
\item {\small $\Code{len(l)} = 1 \land \Code{all\_evens(l)}$}: this covers all
  singleton lists of even elements,
\item
  {\small $\Code{n} \neq 0 \land \Code{len(l)} \neq 1 \land
  \Code{all\_evens(l)}$}: this covers all non-singleton even lists
  where the size parameter is non-zero,
\item {\small
    $(\Code{len(l)} = 1 \lor \Code{empty(l)}) \land
    \Code{all\_evens(l)}$}: this covers even lists with zero or one
  elements.
\end{itemize}

From this solution space, \generalize{} adapts an existing
learning-based specification inference algorithm~\cite{ZDDJ+21} to
find a coverage type that captures the missing outputs of the target
generator:\footnote{The appendices of the full version of the
  paper~\cite{cobb:full} include the full definitions of the
  \generalize{} and \localize{} subroutines, as well as the proof of
  \autoref{thm:generalization+sound}.}

\begin{theorem}[\generalize{} is Sound]
  \label{thm:generalization+sound}
  Given a typing context $\Gamma$, %
  the current coverage $\nuut{b}{\psi_\texttt{cur}}$, %
  and the target coverage $\nuut{b}{\psi}$,
  \generalize($\Gamma, \nuut{b}{\psi_\texttt{cur}}, \nuut{b}{\psi}$)
  produces a $\psi_\texttt{need}$ of the form
  $\bigvee(\bigwedge\overline{\phi} \land \bigwedge \overline{\lnot
    \phi}) \land \psi$ such that
  $\Gamma \covervdash \nuut{b}{\psi_\texttt{cur} \lor
    \psi_\texttt{need}} <: \nuut{b}{\psi}$. Moreover,
  $\psi_\texttt{need}$ is a minimal solution in the solution space
  considered by \generalize{}:%
  \\
  $\lnot \exists\, \psi'{\in}\{ \psi' ~|~ \psi' =
  \bigvee(\bigwedge\overline{\phi} \land \bigwedge \overline{\lnot
    \phi}) \land \psi\}.\; \Gamma \covervdash
  \nuut{b}{\psi_\texttt{cur} \lor \psi'} <: \nuut{b}{\psi} ~~\land~~
  \psi'{\implies}\psi_\texttt{need}$.
\end{theorem} %

\subsection{Localization}
\label{sec:localization}

The \localize{} subroutine inserts holes into a generator $s$, to
produce a sketch, $s'$, and a set of the locations in $s'$,
$\overline{\Gamma_j \vdash \hole{}_j \;:\;\nuut{b}{\phi_j}}$, for the
subsequent \synthesize{} phase to repair. Intuitively, \localize{}
builds $s'$ by inserting holes at the end of every control flow path
in $s$, recording the typing context and missing coverage at that
point.%
\localize{} leaves any existing holes in
$s$ untouched, adding them to the set of repair locations; it also
replaces any \ocamlinline{err}s with holes, as these terms contribute
no useful coverage. \autoref{fig:ex+localize} shows the output of
\localize{} on an incomplete generator for BSTs. Each of the four
holes in the resulting sketch is accompanied by a typing context that
extends the initial context $\Gamma$ with hole-specific control flow
constraints and local variables, e.g., the extended context for
$\hole{}_1$ in the \ocamlinline{then} branch of the generator in
\autoref{fig:ex+localize} is
$\Gamma_1 \equiv \Gamma, \_:\nuot{\Code{bool}}{\Code{n}=0}$.

\begin{figure}[t]
  \begin{tabular}{p{.28\linewidth}|p{.29\linewidth}|p{.36\linewidth}}
    \vspace{-1.15cm}
    \begin{ocaml}[xleftmargin=-6pt, numbersep=0pt, fontsize=\footnotesize]
if n == 0
 then Leaf
 else if lo + 1 < hi then
   let x = int_range lo hi in
    err
  else !\hole{}$_0\; : \;\tau_0$!
   \end{ocaml}
    &
      \vspace{-1.15cm}
    \begin{ocaml}[xleftmargin=0pt, fontsize=\footnotesize]
if n == 0
 then Leaf !$\oplus$! !\hole{}$_1\; : \; \tau_\texttt{need}$ !
 else if lo + 1 < hi then
   let x = int_range lo hi in
    !\hole{}$_2\; : \; \tau_\texttt{need}$!
  else !\hole{}$_0\; :\; \tau_0$! !$\oplus$! !\hole{}$_3\; : \; \tau_\texttt{need}$!
\end{ocaml}
    &
      \footnotesize
      $\left\{~~
      \begin{aligned}
        \Gamma, \noot{\Code{n} = 0} & \vdash \hole{}_1~:~\tau_\texttt{need},\\[3pt]
        \begin{aligned}
          & \Gamma, %
          \noot{\Code{n}\neq 0}, %
          \noot{\Code{lo} +1 < \Code{hi}},  \\[-2pt]
          & ~\Code{x}:\ut{x}{\Code{int}}{\Code{lo} \leq \Code{x} \leq
            \Code{hi}}
      \end{aligned} & \vdash \hole{}_2~:~\tau_\texttt{need}, \\[3pt]
      \Gamma, %
      \noot{\Code{n}\neq 0}, %
      \noot{\Code{hi} \leq \Code{lo} +1}%
      & \vdash \hole{}_0~:~\tau_0, \\[3pt]
      \Gamma, %
      \noot{\Code{n}\neq 0}, %
      \noot{\Code{hi} \leq \Code{lo} +1} %
      & \vdash \hole{}_3~:~\tau_\texttt{need}
    \end{aligned}~~
    \right \}
    $
  \end{tabular}
  \vspace{-.2cm}
  \caption{An incomplete generator for BSTs and the sketch and set of
    holes that \localize{} produces from it, where %
    {\footnotesize
      $\Gamma \equiv \Code{n}{:}\ot{n}{\Code{int}}{\Code{n} \geq 0},
      \, \Code{lo}{:}\Code{int}, \;
      \Code{hi}{:}\ot{hi}{\Code{int}}{\Code{lo}{\leq} \Code{hi}}, \;
      \Code{genBST}{:}\ldots$ }
  We use {\footnotesize $\noot{\phi}$} as shorthand for %
  {\footnotesize $\_~:~\ot{\_}{\Code{bool}}{\phi}$}.}
\label{fig:ex+localize}
\vspace{-.4cm}
\end{figure}

\begin{algorithm}[t!]

  \small \Params{
    $\Gamma$: typing context,%
    $\; s$: A sketch with $j$ holes, %
    $\overline{\Gamma_j \vdash \hole_j~:~\nuut{b}{\psi_j}}$: typing
    contexts and types for the $j$ holes in $s$, %
    $\nuut{b}{\psi}$: target coverage} %
  \Output{A repaired generator $s[\overline{e_j}]$, where
    $\Gamma \vdash s[\overline{e_j}] ~:~
    \nuut{b}{\psi}$} %
  $\overline{Exp_j \leftarrow \emptyset{};}\:$ %
  $\overline{Cand_j \leftarrow \emptyset{};}\:$ %
  $\overline{e_j \leftarrow \Code{err};}$ $\alpha \leftarrow 0$\; %

  \While{$\Gamma \not \vdash s[\overline{e_j}] ~:~ \nuut{b}{\psi}$
    $~~\land~~ \alpha \leq $ \text{\textsc{MaxCost}} }{ %
    \ForEach{$\Gamma_i \vdash \hole_i~:~\nuut{b}{\psi_i} \in
      \overline{\Gamma_j \vdash \hole_j~:~\nuut{b}{\psi_j}}$}{ %
      \lIf
      (\CodeComment{Skip if $\hole{}_i$ has already been repaired})
      {$e_i \neq \Code{err}$}
      {\Continue} %
      \For{$e \in \genExp(Exp_i, \alpha)$} {
        $\tau \leftarrow \typeInferStrict{}(\Gamma_i,\; e)$\;
        \If
        {$\Gamma_i \vdash \tau \equiv \nuut{b}{\bot}$%
          $~\lor~ \exists\, e' \in Exp_i.\; \Gamma_i \vdash e': \tau'
          \land \Gamma_i \vdash \tau' \equiv \tau$}{
          \Continue;
          \CodeComment{Discard $e$ if unsafe or provides no useful
            coverage}
        }
          $Exp_i \leftarrow Exp_i \cup \{e\}$\;
          \If{$\Gamma_i \vdash \nuut{b}{\psi_i} <: \tau$}{
            $Cand_i \leftarrow Cand_i \cup \{e\}$\;
            $Repairs \leftarrow \{e ~|~ \exists Cand \subseteq Cand_i.\; %
            e = \underset{e' \in Cand}{\bigoplus} e' ~~\land~~ %
            \Gamma_i \vdash \typeInferStrict{}(\Gamma_i,\; e) \equiv \nuut{b}{\psi_i} %
            \}$\;
            \lIf(\CodeComment{Use a precise solution for $\hole{}_i$ if one exists})
            {$Repairs \neq \emptyset$}
            {$e_i \leftarrow \text{min}~Repairs$}
          }
      }
    } $\alpha \leftarrow \alpha + 1$\; } %
  \lIf{$\Gamma \vdash s[\overline{e_j}] ~:~\nuut{b}{\psi}$}
  {\Return{$s[\overline{e_j}]$}} %
  \ForEach
  {$\Gamma_i \vdash \hole_i~:~\nuut{b}{\psi_i} \in
    \overline{\Gamma_j \vdash \hole_j~:~\nuut{b}{\psi_j}}$}
  { \lIf
    (\CodeComment{Skip if $\hole{}_i$ already has a precise solution})
    {$e_i \neq \Code{err}$}
    {\Continue} %
    $r \leftarrow \text{min}\,\{ e \in Exp_i ~|~ \typeInferStrict{}(\Gamma_i, e) <: \nuut{b}{\psi_\texttt{i}}\}$;
    \CodeComment{Find term with minimal excess coverage}

    $Repairs \leftarrow \{e ~|~ \exists Exp' \subseteq \{ e' \in Exp_i ~|~ \Gamma_i \vdash r \sqsubset e' \}.\; %
    e = \underset{e' \in Exp'}{\bigoplus} e' ~~\land~~
    \typeInferStrict{}(\Gamma_i, e) <: \nuut{b}{\psi_\texttt{i}} \}$\;
    \leIf(\CodeComment{Use the more precise patch for $\hole{}_i$})
    {$Repairs \neq \emptyset$}%
    {$e_i \leftarrow \text{min}~Repairs$}
    {$e_i \leftarrow r$}
  }
  \Return{$s[\overline{e_j}]$}
\caption{Synthesize repairs (\synthesize{})}
\label{algo:synthesize}
\end{algorithm}

\subsection{Synthesizing Patches}
\label{sec:synthesis}

The final subroutine of our algorithm is \synthesize{}, shown in
\autoref{algo:synthesize}, which generates patches for the holes in
the sketch built by \localize{}. As we saw in \autoref{sec:overview},
while the type produced by \generalize{} provides a top-level goal for
\synthesize{}; this type does not provide guidance on how coverage
should be apportioned to subexpressions, e.g., in the presence of
non-deterministic choice.  For that reason, \synthesize{} uses a
bottom-up approach, adopting an inductive-synthesis-style
algorithm~\cite{Recprogsyn, Burst, Stun} to enumerate a pool of
candidate repairs for each hole.

\synthesize{} maintains two sets of terms for each hole
$\Gamma_i \vdash \hole{}_i~:~\nuut{b}{\psi_i}$: a general pool of all
type-safe terms enumerated so far, $Exp_i$, and a set of candidate
patches $Cand_i$ that cover a portion of the hole's missing
outputs. As discussed in \autoref{sec:overview}, $Cand_i$ is equipped
with a partial order that uses the coverage types of its elements, a
property that \synthesize{} leverages when extracting candidate
patches. \synthesize{} maintains hole-specific sets of terms because
the validity of a repair depends on the context into which it is
inserted: if a patch uses local variables or its safety depends on a
particular set of path conditions, for example. The algorithm also
tracks whether it has found a meaningful repair for each hole, $e_i$;
these are initially set to \Code{err} (line 1). \synthesize{} uses a
loop to iteratively \emph{enumerate} terms below the current cost
$\alpha$ (lines 2-14), \emph{filtering} any enumerated terms that are
not useful (lines 7-8), and then attempts to \emph{extract} a
\emph{precise} patch for each hole from the (partial) solutions it has
found (lines 12-13). This loop terminates when either the current set
of patches are sufficient to ensure the current sketch has the desired
coverage or a limit on the cost of enumerated terms has been reached
(line 2). If a coverage-complete solution is in hand when the loop
ends, it is returned (line 15); this may happen even if some holes
have not been repaired, e.g., if patching a user-provided hole ensures
coverage completeness. If the current set of patches do not make the
sketch coverage complete, \synthesize{} then \emph{extracts} the
\emph{best} patch it can for each unrepaired hole, attempting to
minimize any excess coverage (lines 16-21). We will now describe each
of the four main pieces of \synthesize{} in more detail.

\paragraph{Enumerate}
\synthesize{} uses the \genExp{} subroutine to generate new terms at
the current cost threshold (line 5). \genExp{} is parameterized over a
set of \emph{seeds} and \emph{components} that are used to construct
candidate repairs. The seeds form the initial set of candidate repairs
and includes constants, e.g., \ocamlinline{0}, \ocamlinline{false} and
\ocamlinline{[]}, the default generators for the base types, e.g.,
\ocamlinline{int_gen} and \ocamlinline{genTree}, any variables that
are in scope for a particular hole. The components are used to
construct new terms from previously generated expressions, and
include, e.g., built-in operators, datatype constructors, and
functions. Seeds and components are both equipped with type signatures
characterizing their coverage guarantees. The name of the generator
being repaired is available via a hole's typing context, e.g.,
\ocamlinline{genBST} in \autoref{fig:ex+localize}, enabling patches to
make use of recursive calls. The signature of this recursive component uses a
refinement type to ensure that all recursive calls use strictly
smaller arguments, the signature of \ocamlinline{genEvens}, for
example, is: %
\vspace{-.4cm}\par\nobreak %
{ \footnotesize
  \begin{align*}
    \Code{genEvens} ~~:~~
    \Code{n'}:\ot{n'}{\Code{int}}{n' \geq 0 \land n' < n} \rightarrow%
    \ut{l}{\Code{int~list}}%
    {\lnot \Code{empty(l)} \land \Code{len(l)} \leq \Code{n'} + 1 %
    ~~\land~~ \Code{all\_evens}(l)}%
  \end{align*}
}

\genExp{} is parameterized over a cost function that it uses to
prioritize certain elements in the search space, a common strategy in
the program synthesis literature~\cite{Gulwani+10, Gulwani+11,
  Recprogsyn}. Cost functions are required to be monotonic--- a term
never has a smaller cost than any of its subterms--- and stateless---
a cost of a term is independent of the terms \genExp{} has already
seen. The cost function used in our implementation of \synthesize{}
prefers terms with the following properties:
\begin{description}[leftmargin=\parindent]
\item[Recursive Calls] We assign a low cost to recursive calls, as
  they typically %
  provide a large amount of coverage, tailored to the current repair.
\item[Same Type as Target] A valid patch must produce values of the
  target coverage type, so we prioritize components that construct
  expressions with the same base type as the target over those that do
  not.
\item[Seed Generators] Default type generators like
  \ocamlinline{int_gen()} often provide useful coverage and are
  prioritized, alongside variables, over constant expressions.
\item[Diverse Terms] A na\"ive enumeration strategy will produce many
  terms that have repeated uses of the same components and seeds,
  e.g. \ocamlinline{Cons(x, Cons(y, []))}. Recursive calls often
  produce these sorts of terms in a more general way, so we prioritize
  expressions comprised of diverse subterms.
\end{description}

\paragraph{Filter}
A key challenge in enumerative approaches to program synthesis is
keeping the set of generated terms from becoming intractably
large. Accordingly, \synthesize{} curates its pool of terms by
discarding expressions that are redundant or unlikely to contribute to
a solution (lines 6-8). While \genExp{} employs a purely
\emph{syntactic} cost function to prioritize terms, \synthesize{} uses
the \emph{semantics} of a term when deciding whether it should be
pruned. This semantic information is encoded in the coverage type
inferred by \typeInferStrict{} (line 6). \typeInferStrict{} is a more
restrictive version of \typeInfer{} which aggressively rejects
function applications in order to filter potentially unsafe
terms. \typeInferStrict{} does so by strictly limiting where type
subsumption may be applied, so that the inferred coverage types of any
function arguments do not include values that violate the refinement
types of its parameters. \typeInferStrict{} implements the following
modified version of the typing rule for function applications: %
\vspace{-.4cm}\par\nobreak %
{\small
\begin{align*}
  \begin{prooftree}
    \hypo{
      \begin{matrix}
        \Gamma \vdash v_1 \typeinfer a: \ot{\nu}{b}{\phi_1} \rightarrow \tau_1
        \quad \Gamma \vdash v_2 \typeinfer c: \ut{\nu}{b}{\phi_2}
        \quad \disjop (\neg (\phi_1), \phi_2) = \emptyset \\
        \quad \Gamma' = a:\ut{\nu}{b}{\nu = c \land \phi_1}, x:\tau_1
        \quad \Gamma, \Gamma' \vdash e \typeinfer \tau
        \quad \tau' = \existsop(\Gamma', \tau)
        \quad \Gamma \vdashunder \tau'
      \end{matrix}
    }
    \infer1[\textsc{SynAppBase'}]{
      \Gamma \vdash \zlet{x}{v_1\ v_2}{e} \typeinfer \tau'
    }
  \end{prooftree}
\end{align*}}
\noindent
This stronger rule ensures, e.g., that \synthesize{}'s pools
of terms only include recursive calls whose size argument is strictly
decreasing.

If \typeInferStrict{} successfully infers a type for a term
$\Gamma_k \vdash e ~:~ \tau$, \synthesize{} then decides whether to
include $e$ in $Exp_k$, using $\tau$ to judge whether it supplies any
meaningful new coverage (line 7). It prunes any $e$ that may not
produce any values at all, i.e., when $\tau$ is equivalent to
$\nuut{b}{\bot}$. \synthesize{} also discards $e$ if it has already
enumerated a \emph{coverage equivalent} term: if two terms cover the
same outputs, we only need to keep around the cheaper one, similarly
to how other synthesizers use observational equivalence to prune
enumerated terms~\cite{Recprogsyn, Duet}. Any terms that are not
filtered are then added to $Exp_k$ (line 9); finally, \synthesize{}
additionally checks whether $e$ provides part of the target coverage,
adding it to the poset of partial solutions if so (line 11).

\paragraph{Extract a Precise Repair}
After enumerating all the useful terms for the current cost bound,
\synthesize{} attempts to extract a \emph{precise} patch, i.e., one
that supplies exactly the coverage required by the current hole
$\hole{}_i$, by using $\oplus$ to combine the partial solutions in
$Cand_i$ (lines 12-13). In order to do efficiently, our implementation
of \synthesize{} leverages the fact that $Cand_i$ is a join
semi-lattice ordered using the inferred types of its elements.
The insight is that we can efficiently check if $Cand_i$ contains a
complete repair by only examining the elements that are direct
supertypes of the target coverage type for a hole $\hole{}_i$, since:
\vspace{-.4cm}\par\nobreak %
\begin{align*}
  \Gamma_i \vdash \underset{e \in Cand'}{\bigoplus} e :
  \nuut{b}{\psi_\texttt{need}} \Rightarrow %
  \Gamma_i \vdash \underset{e \in Cand_i}{\bigoplus} e :
  \nuut{b}{\psi_\texttt{need}}
\end{align*}
\vspace{-.2cm}
where
\vspace{-.4cm}\par\nobreak %
\begin{align*}
  Cand' \equiv
  &
    \left\{\begin{tabular}{r|l}
             \multirow{2}{*}{$e \in Cand_i$} %
             & $\phantom{\land~}
               \nuut{b}{\psi_\texttt{need}} <: \typeInferStrict{}(\Gamma_i, e) $ \\
             & $\land~ \not \exists e' \in Cand_i.\; ~%
               \nuut{b}{\psi_\texttt{need}} <: \typeInferStrict{}(\Gamma_i, e') <: \typeInferStrict{}(\Gamma_i, e)$%
           \end{tabular}\right\}
\end{align*}

\begin{example}
  To see how we leverage this to extract a patch from a poset of
  candidate solutions, we will show how \synthesize{} builds a precise
  solution for the hole {\small
    $\Code{n}:\ot{n}{\Code{int}}{\Code{n} \geq 0} \vdash \hole{} ~:~
    \ut{l}{\Code{int~list}}{\Code{len(l)} \leq \Code{n}}$}. We begin
  immediately after the term \Code{e$_5$} has been added to the
  lattice at the top of \autoref{fig:ex+refine}.  We first see if
  \sCode{e$_5$} itself is a precise patch by checking if its inferred
  type is equivalent to the target coverage
  $\ut{l}{\Code{int~list}}{\Code{len(l)} \leq \Code{n}}$. Since this
  fails, we instead temporarily insert a ``dummy'' node with the
  target type into the lattice, producing the lattice at the bottom of
  \autoref{fig:ex+refine}.  Observe that the inserted node is a direct
  parent of \sCode{e$_3$}, \sCode{e$_4$}, and \sCode{e$_5$}, and that
  furthermore the join of these three expressions has the coverage
  type %
  {\small
    $\ut{l}{\Code{int~list}}{\Code{empty(l)} \lor \Code{len(l)} =
      \Code{n} \lor \Code{len(l)} < \Code{n}}$}. Since this type is
  equivalent to the type of the target hole, we have found a valid
  patch, which we adapt as the solution for this hole.\footnote{Note
    that this extraction strategy crucially depends on using the
    precise type inferred by \typeInferStrict{} to order the elements in
    the lattice. Using the subtyping relation on any valid type (via
    subsumption) would allow an element's children to provide more
    coverage than it does: under such a strategy, \Code{int\_gen()}
    could be a child of any element, for example!}
\end{example}

\begin{figure}[!t]
  \centering
  \begin{tikzpicture}[->,>=stealth',level 1/.style={sibling distance = .35\linewidth},
    level 2/.style={sibling distance = .5\linewidth}]
    \node  { \footnotesize %
      $\Gamma\,\vdash$ \Code{e$_1$} $\, : \, \ut{l}{il}{\top}$
    }
    child{ node[yshift=15pt] {
        \footnotesize
        $\Gamma\,\vdash$ \Code{e$_2$} $\, : \,$ $\ut{l}{il}{
          \lnot \Code{empty(l)}}$%
      }
      child{ node[yshift=15pt]  {
            \footnotesize
           $\Gamma\,\vdash$ \Code{e$_3$} $\, : \,$
           $\ut{l}{il}{\lnot \Code{empty(l)}  \land \Code{len(l)} = \Code{n} }$%
        }
        edge from parent [->] node [above, rotate=35] {$\sqsupseteq$}
      }
      child{ node[yshift=15pt]  {
            \footnotesize
            $\Gamma\,\vdash$ \Code{e$_4$} $\, : \,$
            $\ut{l}{il}{\lnot \Code{empty(l)} \land \Code{len(l)} < \Code{n}}$%
      }
      edge from parent [->] node [above, rotate=-45] {$\sqsubseteq$}
      }
      edge from parent [->] node [above, rotate=30] {$\sqsupseteq$}
    }
    child{ node[yshift=15pt]  {
          \footnotesize
          $\Gamma\,\vdash$ \Code{e$_5$} $\, : \,$
          $\ut{l}{il}{\Code{empty(l)}}$%
      }
      edge from parent [->] node [above, rotate=-30] {$\sqsubseteq$}
    }
    ;
\end{tikzpicture}
\noindent\rule{.8\textwidth}{.5pt}
  \begin{tikzpicture}[->,>=stealth',level 1/.style={sibling distance = .45\linewidth},
    level 2/.style={sibling distance = .37\linewidth}]
    \node  { \footnotesize %
       $\Gamma\,\vdash$ \Code{e$_1$} $\, : \, \ut{l}{il}{\top}$
    }
    child{ node[yshift=15pt] {
        \footnotesize
        $\Gamma\,\vdash$ \Code{e$_2$} $\, : \,$ $\ut{l}{il}{
          \lnot \Code{empty(l)}}$%
      }
      child{ node[yshift=15pt] (e3) {
            \footnotesize
          $\Gamma\,\vdash$ \Code{e$_3$} $\, : \,$
          $\ut{l}{il}{\lnot \Code{empty(l)} \land
            \Code{len(l)} = \Code{n} }$%
        }
        edge from parent [->] node [above, rotate=35] {$\sqsupseteq$}
      }
      child{ node[yshift=15pt] (e4)  {
            \footnotesize
            $\Gamma\,\vdash$ \Code{e$_4$} $\, : \,$
            $\ut{l}{il}{\lnot \Code{empty(l)} \land \Code{len(l)} < \Code{n} }$%
        }
        edge from parent [->] node [above, rotate=-45] {$\sqsubseteq$}
      }
      edge from parent [->] node [above, rotate=30] {$\sqsupseteq$}
    }
    child{ node[yshift=15pt] (target) {
          \footnotesize
          \textcolor{CBGreen}{$\Gamma\,\vdash$ \Code{target} $\, : \,$
            $\ut{l}{il}{\Code{len(l)} \leq \Code{n}}$}%
      }
      child{ node[xshift=20pt, yshift=15pt]  {
          \footnotesize
          $\Gamma\,\vdash$ \Code{e$_5$} $\, : \,$
          $\ut{l}{il}{\Code{empty(l)}}$%
        }
        edge from parent [->] node [above, rotate=-30] {$\sqsubseteq$}
      }
      edge from parent [->] node [above, rotate=-30] {$\sqsubseteq$}
      (target) edge  [->] node [above, rotate=15] {$\sqsupseteq$} (e4)
      (target) edge  [->] node [above right, rotate=15] {$\sqsupseteq$} (e3)
    }
  ;
\end{tikzpicture}
\vspace{-.2cm}
\caption{A lattice of repairs before and after
  inserting \textcolor{CBGreen}{target}, where
  $\Gamma \equiv \Code{n}:\ot{n}{\Code{int}}{\Code{n} \geq 0}$ and $il
  \equiv$ \ocamlinline{int list}}
\label{fig:ex+refine}
\vspace{-.5cm}
\end{figure}

\paragraph{Extract an Imprecise Repair}
If \synthesize{} hits its cost bound without finding precise solutions
that complete the input sketch, it then extracts the best
\emph{imprecise} patch it can for each unrepaired hole (lines
16-20). This process proceeds similarly to the one for precise
repairs, with a few key tweaks. Since we know we will not find a
precise solution for the current hole, we identify the element from
the complete pool of enumerated terms that has the minimal excess
coverage (line 18). This term corresponds to the parent of a
(hypothetical) element with the target type in our lattice; we
identify it by once again inserting a ``dummy'' node into
$Exp_i$.\footnote{In the worst case, this will be the default
  generator for the target type, as this is always included in our
  set of seeds.}  \synthesize{} then tries to refine this solution by
combining subsets of its children in the lattice using $\oplus$ (line
19); \synthesize{} uses the smallest such combination with sufficient
coverage, if one exists (line 20).

\begin{theorem}[\complete{} is Sound and Total]
  \label{thm:complete+sound}
  Given a program $s$ that is well typed under typing context
  $\Gamma$, $\Gamma \covervdash s \; : \; b$, and target coverage type
  $\nuut{b}{\psi}$, \complete{} returns a coverage complete generator,
  $\Gamma
  \covervdash \complete(s,~\Gamma,~\nuut{b}{\psi}) \; : \; \nuut{b}{\psi}$.
\end{theorem}

\section{Implementation}
\label{sec:implementation}

\name{}, our prototype implementation of the above approach, consists
of about 3k lines of OCaml~\cite{leroy2024ocaml}, and uses a modified
version of \poirot{}~\cite{poirot} as its coverage type checker; this
type checker uses Z3~\cite{de2008z3} as its backing SMT
solver. \name{} ingests and outputs sized generators in a DSL that
closely mimics \langname{}. We have implemented this language as a
shallowly embedded DSL in OCaml, and repaired generators can be
directly executed using OCaml's QCheck
framework~\cite{QCheck}. \name{} is parameterized over a set of base
types, components, and method predicates. It currently supports a
number of standard OCaml primitive operations and datatypes, and
includes built-in method predicates for expressing properties of these
types, e.g., \ocamlinline{empty} and \ocamlinline{sorted}.

The guarantees provided by the \complete{} algorithm are
\emph{possibilistic}: the coverage guarantees of weighted and fair
implementations of $\oplus$ are the same. In practice, however, users
often prefer generators that bias the choices of $\oplus$. If only one
of its choices includes a recursive call, for example, a fair
implementation of $\oplus$ will bias a generator towards smaller
values: \ocamlinline{genTree} in the introduction generates
\ocamlinline{Leaf} nodes half of the time, for example. Thus, \name{}
adopts the commonly used approach of using the bound of a sized
generator to bias uses of $\oplus$ operators~\cite{quickchick}: after
synthesis, \name{} applies a syntactic transformation to adjust the
weights of $\oplus$ in which only a single choice has a recursive
call, weighting that choice according to the current bound. In
practice, this means that \name{} produces generators that are
initially biased towards recursive calls, but which are more likely to
take the other choice as \ocamlinline{size} decreases.

\section{Evaluation}
\label{sec:evaluation}

Using \name{}, we have investigated five key questions about our
approach to generator repair:

\begin{itemize}
\item[{\bf RQ1}] Is our approach able to automatically find complete
  repairs for different kinds of generators, covering a diverse set of
  properties and datatypes, in a reasonable amount of time?
\item[{\bf RQ2}] Is \name{} effective when used as a sketch-based
  synthesizer? Can it produce a generator from a skeleton that
  contains only the desired control flow structure of the target
  generator?
\item[{\bf RQ3}] How does our approach compare to alternative repair
  strategies that exclusively prioritize either safety or
  completeness?
\item[{\bf RQ4}] How sensitive is our approach to the set of
  components used?
\item[{\bf RQ5}] How do our statically repaired generators compare to
  alternative complete input generation approaches that rely on
  run-time constraint solving?
\end{itemize}

\noindent All of our experiments were run on a 2020 M1 13-inch MacBook
Pro with 8 GB of memory.

\subsection{Synthesis of Coverage Complete Generators (RQ1, RQ2)}
\label{sec:RQ1}

\begin{table}[t!]
\renewcommand{\arraystretch}{0.8}
\caption{\small The results of using \name{} to repair incomplete
  generators. Benchmarks are annotated with their source:
  QuickChick~\cite{quickchick} ($^*$), \citet{LHP+19} ($^\star$) and
  \citet{ZDDJ+21} ($^\diamond$). The middle set of columns
  characterize the complexity of the problem and the solution: the
  number of holes in the initial sketch (\#Holes) and the size of the
  AST of the term that is synthesized (Repair Size). The last set of
  columns describe the effort required to find a repair: the number of
  terms enumerated (\#Terms), the number of SMT queries (\#Queries),
  the time it took to infer the missing coverage (Abduction), the time
  spent generating the final solution (Synthesis), and the total time
  needed to find a coverage complete generator (Total). }
\label{tab:eval+RQ1}
\vspace*{-.05in} \footnotesize
\begin{tabular}{r|rr|rrrrr}
\midrule
                Benchmark &   \#Holes &   Repair Size &   \#Queries &   \#Terms &   Abduction (s) &   Synthesis (s) &   Total Time(s) \\
\midrule
         Sized List$^*$ 1 &        1 &             1 &         31 &        3 &            0.18 &            0.4  &            0.61 \\
                        2 &        1 &             1 &         30 &        3 &            0.15 &            0.32 &            0.51 \\
                        3 &        1 &            14 &        105 &       22 &            0.18 &            1.67 &            1.89 \\
                        4 &        2 &             2 &         38 &        6 &            0.2  &            0.39 &            0.63 \\
                        5 &        1 &            20 &        122 &       22 &            0.21 &            2.02 &            2.27 \\
                        6 &        2 &            15 &        112 &       25 &            0.17 &            1.7  &            1.9  \\
                        7 &        2 &            21 &        128 &       25 &            0.3  &            2.04 &            2.38 \\
                        8 &        1 &            20 &        121 &       22 &            0.19 &            2.14 &            2.37 \\
                   sketch &        2 &            21 &        127 &       25 &            0.37 &            2.03 &            2.43 \\
\midrule
     Duplicate List$^*$ 1 &        1 &             1 &         36 &        4 &            0.11 &            0.66 &            0.8  \\
                        2 &        1 &            18 &        110 &       67 &            0.07 &            4.09 &            4.19 \\
                   sketch &        2 &            19 &        121 &       71 &            0.17 &            4.48 &            4.68 \\
\midrule
 Unique List$^\Diamond$ 1 &        1 &             1 &         30 &        3 &            0.1  &            0.29 &            0.43 \\
                        2 &        2 &             7 &         68 &       16 &            0.08 &            1.08 &            1.18 \\
                   sketch &        3 &             8 &         71 &       19 &            0.11 &            1.09 &            1.22 \\
\midrule
        Sorted List$^*$ 1 &        1 &             1 &         35 &        4 &            0.12 &            0.36 &            0.51 \\
                        2 &        2 &            16 &        291 &      226 &            0.09 &            4.47 &            4.59 \\
                   sketch &        3 &            17 &        298 &      230 &            0.11 &            4.46 &            4.61 \\
\midrule
          Even List 1 &        1 &            11 &        123 &       35 &            0.38 &            1.91 &            2.34 \\
                        2 &        1 &            11 &        199 &       48 &            0.33 &            4.76 &            5.14 \\
                        3 &        1 &            17 &        202 &       58 &            0.38 &            5.42 &            5.84 \\
                        4 &        2 &            22 &        301 &       83 &            0.48 &            6.3  &            6.82 \\
                        5 &        1 &            33 &        261 &       58 &            0.48 &            7.42 &            7.98 \\
                        6 &        2 &            28 &        304 &       93 &            0.51 &            7.06 &            7.61 \\
                   sketch &        2 &            40 &        355 &       93 &            0.65 &            8.86 &            9.56 \\
  \midrule\midrule
         Sized Tree$^*$ 1 &        1 &             1 &         37 &        3 &            0.36 &            0.56 &            0.99 \\
                        2 &        1 &             1 &         36 &        3 &            0.32 &            0.55 &            0.95 \\
                        3 &        1 &            18 &        165 &       17 &            0.36 &            6.31 &            6.75 \\
                   sketch &        2 &            25 &        188 &       20 &            0.5  &            6.88 &            7.45 \\
\midrule
  Complete Tree$^\star$ 1 &        1 &             1 &         34 &        3 &            0.18 &            0.36 &            0.6  \\
                        2 &        1 &            18 &        218 &       45 &            0.17 &            3.39 &            3.65 \\
                   sketch &        2 &            19 &        225 &       48 &            0.43 &            3.3  &            3.78 \\
\midrule
            BST$^\star$ 1 &        1 &             1 &         78 &        5 &           49.06 &            2.73 &           56.21 \\
                        2 &        1 &             1 &         85 &        6 &           49.28 &            3.27 &           56.88 \\
                        3 &        1 &             1 &         77 &        5 &           47.92 &            3.13 &           55.17 \\
                        4 &        1 &            33 &      13312 &    11136 &           42.72 &          591.28 &          637.67 \\
                   sketch &        3 &            57 &      13401 &    11146 &          169.45 &          595.26 &          765.49 \\
  \midrule\midrule
     \vspace{6pt} Red-Black Tree$^*$ 1 &        1 &             1 &        156 &        5 &           33.16 &            2.03 &           35.85 \\
                        2 &        1 &             1 &        154 &        6 &           32.3  &            2.07 &           35.19 \\
                        3 &        1 &            14 &        232 &       45 &           30.23 &            3.92 &           35.14 \\
                        4 &        1 &            34 &        362 &       95 &           32.52 &            8.67 &           41.95 \\
                        5 &        1 &            31 &        328 &       87 &           29.92 &            7.1  &           38.32 \\
                        6 &        1 &            17 &        219 &       51 &           30.58 &            2.72 &           33.68 \\
                   sketch &        4 &           108 &        673 &      232 &           45.59 &           16.53 &           62.58 \\
\hline
\end{tabular}
\vspace{-.5cm}

\end{table}

Our first set of experiments evaluate the ability of \name{} to
automatically repair an incomplete input generator, and considers a
diverse set of data types (e.g., lists, trees, and lambda terms) and
target preconditions (e.g., sorted lists, balanced trees, and
well-typed lambda terms) (\textbf{RQ1}). To build the incomplete
generators used in our experiments, we took coverage-complete
generators drawn from the existing PBT literature~\cite{quickchick,
  LHP+19, ZDDJ+21}, and made them incomplete by replacing one or more
of their branches with \ocamlinline{err}. We construct multiple
variants of each generator by removing different combinations of
branches, including \emph{sketches} of each generator which replace
all its branches with \ocamlinline{err} (\textbf{RQ2}). The coverage
type specification used in each benchmark is a direct translation of
the precondition of the function under test. \autoref{tab:eval+RQ1}
presents the results of using \name{} to repair each of these
variants. The variants are (roughly) ordered by the amount of the
functionality they lack, with the sketch acting as the final variant
of each generator. The required repairs range from the relatively
trivial--- the first two sized list variants only require inserting an
empty list (\ocamlinline{[ ]}), for example--- to the more
substantial: repairing the red-black tree sketch requires synthesizing
multiple recursive calls and applications of datatype constructors
with very specific arguments. Each benchmark uses one of three sets of
components based on its base type, i.e., list, tree, or red-black
tree. Each set includes all the components needed by at least one of
the original generators targeting that base type.

\begin{wrapfigure}{r}{.48\linewidth}
  \vspace{-.35cm}
  \begin{tabular}{l}
    \begin{minipage}{.9\linewidth}
  \begin{ocaml}[fontsize = \footnotesize, xleftmargin=0pt]
let lt = rbtree_gen (inv - 2) false (h - 1) in
let rt = rbtree_gen (inv - 2) false (h - 1) in
Rbtnode (false, lt, int_gen(), rt)
\end{ocaml}
\end{minipage}
    \\[13pt] \hline
    \\[-10pt]
    \begin{minipage}{.9\linewidth}
  \begin{ocaml}[fontsize = \footnotesize, xleftmargin=0pt]
rbtree_gen (inv - 1) true h
  \end{ocaml}
  \end{minipage}
  \end{tabular}
  \vspace{-8pt}
  \caption{The relevant part of the original and
    \name{}-repaired versions of the sixth red-black tree
    variant.}
  \label{fig:rb+example}
  \vspace{-8pt}
\end{wrapfigure}
For almost every variant, \name{} was able to find a repair that was
equivalent to the term that had been replaced by \Code{err}, modulo
some syntactic differences (e.g., order of operations, normal form),
including for every sketch (\textbf{RQ2}). A notable exception is the
sixth variant of the red-black tree generator, shown in
\autoref{fig:rb+example}: while the original generator directly
constructs a black node and its subtrees, \name{} finds a smaller, but
semantically equivalent repair that makes a recursive call with the
correct color/invariant arguments. Our cost function biases recursive
terms earlier in the synthesis process because they produce similar
coverage to that of our goal. In this case, the coverage supplied by
the original branch can be fully realized by flipping the color in the
recursive call and updating the size invariant.

For all these benchmarks, \name{} was able to find a complete
generator within a reasonable amount of time (\textbf{RQ1}), with the
time taken roughly correlated to the complexity of the target property
and the functionality that was removed. Most of the repair time is
spent on calls to Z3, with \generalize{} and \synthesize{} dominating
the total runtime. Longer abduction times generally correspond to a
more complex specification and more method predicates, while longer
synthesis times correspond to a larger space of candidate patches. As
expected, repairing the sketch of each benchmark takes the most time,
as they are missing the most coverage. The BST sketch, for example,
requires \name{} to explore one of the largest search spaces of all
our experiments, with the final repair synthesizing a pair of
recursive calls with non-trivial arguments. On the other end of the
spectrum, the target coverage type of the red-black tree benchmarks
enforces several non-trivial invariants, resulting in some of the
longest abduction times. Most of the remainder of the total time is
spent type checking the completed generator; these times are
consistent with those reported by \citet{poirot}.

\subsubsection*{Case Study: Simply-Typed Lambda Calculus Terms}

As a final experiment, we also investigated \name{}'s performance on a
more challenging problem: repairing a generator for well-typed simply
typed lambda calculus (STLC) terms~\cite{LGHHPX+17, PCRH+11}. %
On its own, the reference generator is already quite complex,
featuring multiple inductive datatypes and auxiliary
functions~\cite{poirot}. The specifications of the generator and these
auxiliary functions are similarly intricate, requiring 15 method
predicates. We developed three variants of the reference generator
using the same methodology as our previous set of experiments. We
additionally bound the space of candidate repairs in each experiment,
by manually limiting the set of components used by \name{} to those
occurring in the expression that was deleted from the reference
implementation. \autoref{tab:eval+stlc} shows the results of this
experiment.

Despite the challenges inherent in this benchmark, \name{} was able to
find complete repairs for all three STLC variants, with the two
simpler variants each taking less than two and a half minutes to
repair. %
The final variant required a more substantial repair that involved
multiple recursive calls and sophisticated reasoning, e.g., the patch
must randomly divide the maximum number of applications allowed in
recursively generated subterms. While searching for this patch,
\name{} enumerates more than 1500 terms and issues almost 4500 SMT
queries. While \name{} is able to successfully find this patch, it
takes almost 10 minutes to do so, with the bulk of the time being
spent querying Z3. We suspect that \name{} will need further
optimizations to scale up to larger sets of components and to mitigate
the inherent complexity of the verification conditions handed off to
Z3.

\begin{table}[t!]
  \renewcommand{\arraystretch}{0.8}
  \caption{Results for the Simply Typed Lambda Calculus case
    study. The columns are the same as \autoref{tab:eval+RQ1}.}
    \label{tab:eval+stlc}
    \centering \vspace*{-.05in} \footnotesize
\begin{tabular}{r|rr|rrrrr}
\hline
              Benchmark &   \#Holes &   Repair Size &   \#Queries &   \#Terms &   Abduction (s) &   Synthesis (s) &   Total Time(s) \\
\hline
  STLC 1 &        1 &             9 &         99 &        9 &            1.81 &            5.55 &            8.29 \\
  2 &        1 &            30 &        694 &      276 &            0.76 &           83.76 &           86.34 \\
  3 &        1 &            59 &       4460 &     1555 &            0.61 &          610.18 &          611.49 \\
\hline
\end{tabular}
\end{table}

\subsubsection*{Discussion}
Taken together, these two sets of experiments provide evidence that
\name{}'s runtime scales reasonably well with the complexity of both
the repair and synthesis tasks, suggesting the potential of our
approach in future applications that depend on generating data that
meets some desired property. Importantly, the cost of performing a
repair is paid once: a repaired generator can be run normally, without
any need to invoke an SMT solver.

\subsection{Comparison with Alternative Repair Strategies (RQ3)}
\label{sec:eval+repair}
At a high-level, \name{} balances two competing concerns when
searching for a patch, trying to find a repair that limits the number
of `useless' inputs that fail to meet the target precondition, while
simultaneously ensuring it does not omit any `interesting' values that
do. This set of experiments compares \name{} to alternative strategies
that exclusively prioritize one of these concerns (\textbf{RQ3}).

\paragraph{Completeness-Focused Repair}
Our first set of experiments compares \name{} against an approach that
only prioritizes completeness when searching for repairs. This admits
an easy implementation: we simply fill in each hole inserted by
\localize{} with the default generator for the base type of the hole,
e.g., \ocamlinline{genTree} or \ocamlinline{int_gen}. This results in
generators that are at least as complete as those found by \name{}, at
the cost of potentially producing more 'useless' inputs. Thus, to
compare the two strategies, we track how many values a repaired
generator produces that violate the target precondition.
We use each generator to produce 20k values, recording how many of
these outputs satisfy the target precondition. Following prior work,
we constrain the size parameter used in each experiment, adopting
similar bounds to those works~\cite{Target,LGHHPX+17,quickchick}; the
discussion in \autoref{sec:limitations} provides more details on the
bounds used. These experiments also address the feasibility of
directly using a default generator to compensate for a generator's
missing coverage.

\begin{figure}[t]
  \centering
  \includegraphics[scale=.26]{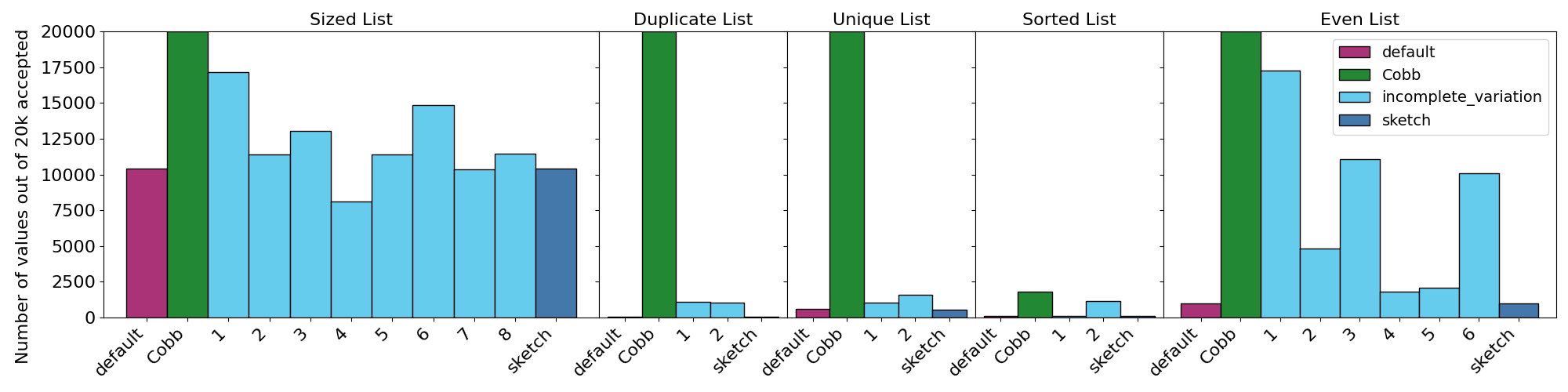}
  \vspace{5pt}
  \includegraphics[scale=.26]{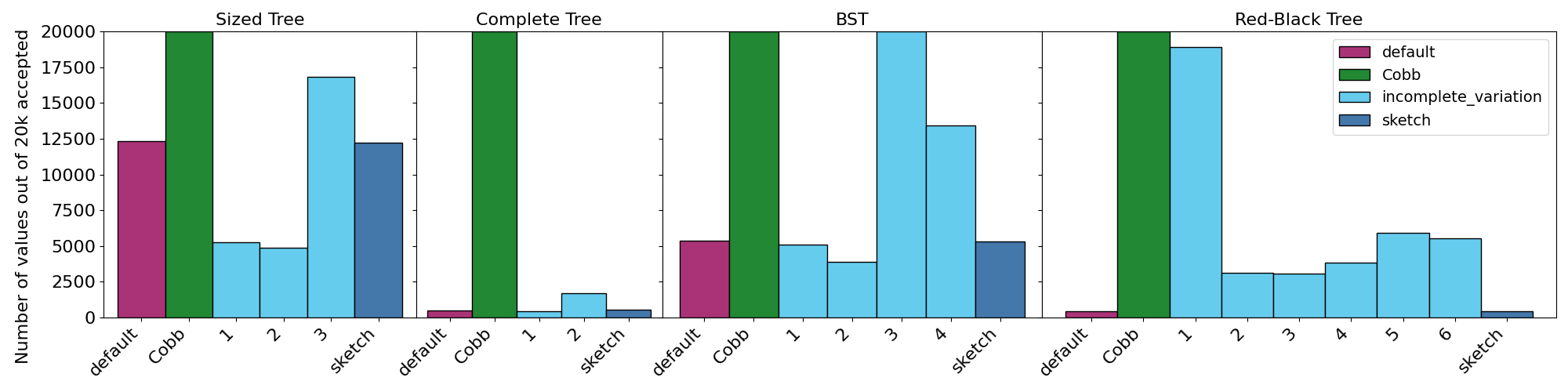}
  \vspace{-10pt}
  \caption{Number of values satisfying the target precondition
    produced by the default, \name{}-repaired, and
    completeness-focused-repaired generators for the sketches in
    \autoref{tab:eval+RQ1}.}
  \label{fig:eval+coverage+list+tree}
  \vspace{-10pt}
\end{figure}

\autoref{fig:eval+coverage+list+tree} presents the results of this
experiment for each of the list and tree benchmarks from
\autoref{tab:eval+RQ1}, respectively. Each graph also reports the
number of valid outputs produced by a default generator; this serves
as a rough proxy for the restrictiveness of the target
precondition. From left to right, each group of columns in the figures
report the number of valid inputs produced by the default generator
(\textcolor{CBPurple}{purple}), the generator repaired by \name{}
(\textcolor{CBGreen}{green}),\footnote{\autoref{fig:eval+coverage+list+tree}
  uses the generator produced from the sketch for the generator
  produced by \name{}. Since the \name{}-repaired generators are
  semantically equivalent, so are their results on these benchmarks.}
and the repaired version of each variant in \autoref{tab:eval+RQ1}
produced by a completeness-focused repair strategy
(\textcolor{CBCyan}{cyan}), ending with the repaired sketch
(\textcolor{CBBlue}{blue}). Unsurprisingly, the last variant performs
comparably to the default generator: applying the completeness-focused
repair strategy to \ocamlinline{genEven}$_\texttt{inc}$, for example,
results in a function that is effectively equivalent to the default
\ocamlinline{int list} generator.

While the generators produced by \name{} are consistently more likely
to produce valid values than their completeness-focused counterparts,
\autoref{fig:eval+coverage+list+tree} shows that the latter strategy can be
effective in certain cases, especially when pitted against the default
generator. As expected, one such scenario is when the target property
is relatively permissive, as is the case for our sized list and sized
tree benchmarks. These generators only need to produce a value within
the expected size; as the default generators show, roughly half of all
randomly generated values satisfy this property.

Conversely, when the target specification is more restrictive, the
alternative repair strategy is less effective. The complete tree
benchmark falls into this category, as the subtrees of a randomly
generated tree are unlikely to have a uniform depth. Similarly, the
target preconditions used by our unique and duplicate list benchmarks
are considerably tighter than that of the sized list benchmark: both
require a list containing \emph{exactly} \ocamlinline{size}
elements. In both cases, the coverage provided by the repaired
generators produced by the alternative strategy is mostly limited to
when the size parameter is small, although uniqueness of list elements
is a slightly more forgiving property.

A similar phenomenon occurs in the BST and red-black tree benchmarks,
albeit in a more nuanced way. Both of these benchmarks feature
semantically rich specifications, while still being somewhat more
permissive than the previous three examples. Notably, the
completeness-focused repair strategy is effective for some of these
benchmarks, in particular the third BST variant and the first
red-black tree variant. For these two examples, the required repairs
fall into execution paths which are exercised very rarely, so the
default generator used in the repair is not given many opportunities
to inject an invalid value into the output of the repaired
generator. In the case of the third BST variant, for example, the
repair is inserted into a branch in which bounds force the BST to be
empty, i.e., \(\Code{lo} + 1 = \Code{hi}\), a scenario that depends on a very
particular sequence of nondeterministic choices. In the case of the
red-black tree, the patch is similarly only executed when the
generator is called with very specific values, namely when the input
black height is precisely zero and the color argument is black. These
sorts of corner cases are sometimes explicitly listed in handwritten
generators in order to improve the likelihood they will occur;
identifying and prioritizing these sorts of corner cases in repairs is
an interesting direction for future work.

On most of these benchmarks, the generators repaired by \name{} almost
always produce valid inputs, with the sorted list generator being the
notable exception.  %
This generator is unique among our benchmarks as it is the only
benchmark in which the reference generator includes an \Code{err}
expression that is hit with some frequency. While errors are fine from
the perspective of our coverage type system--- the right sequence of
nondeterministic choices always allows the generator to avoid them---
because the original sorted list generator does not implement any kind
of backtracking~\cite{LGHHPX+17}, these errors impact the number of
sorted lists the repaired generator produces. As a result, the
repaired generators in this experiment only have a reasonable
probability of yielding a valid output for smaller lists.

\begin{figure}[t]
  \centering
  \includegraphics[scale=.26]{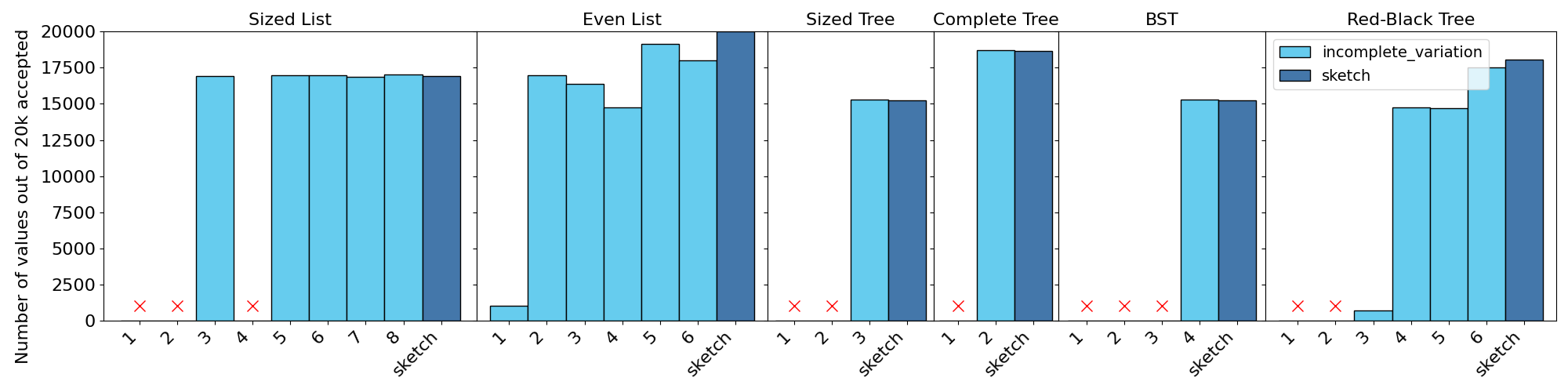}
  \vspace{-10pt}
  \caption{Number of tests produced by \name{}-repaired generators
    from \autoref{tab:eval+RQ1} that fall outside the range of
    repaired versions that used a safety-focused strategy. The
    safety-repaired versions of the three omitted generators were
    coverage complete.}
  \label{fig:eval+safety}
  \vspace{-10pt}
\end{figure}

\paragraph{Safety-Focused Repair}
As the previous experiment showed, a repair strategy that only
prioritizes completeness is likely to produce generators that output
\emph{useless}, i.e., invalid values. This set of experiments
investigates whether a repair strategy that only considers safety will
yield generators that are likely to omit \emph{interesting}, i.e.,
valid values. Before discussing our results, observe that it is not
obvious how to measure the incompleteness of a generator: even an
incomplete generator can produce an infinite number of valid
inputs. Our strategy is to instead compare the relative completeness
of two generators, in this case by quantifying the number of outputs
produced by a \name{}-repaired generator that could never be produced
by a generator repaired using a safety-focused strategy. In detail, we first
ascribe a standard refinement type to the return type of each safe
generator: intuitively, the qualifier of this type overapproximates
the range of the generator. Negating this qualifier characterizes the
set of outputs that fall outside the range of a safety-repaired
generator; any valid outputs of a \name{}-repaired generator that
satisfy this negated property could never have been produced by its
safe counterpart.

As an example, one way to repair \ocamlinline{genIntList} from
\autoref{fig:sized+list+impl} so that it is safe with respect to our
target precondition is:
\begin{ocaml}
  let rec genIntList!$_\texttt{safe}$! (n : int) : int list = if n == 0 then [ ] else [ ]
\end{ocaml}
\noindent We can ascribe the following refinement type to
\ocamlinline{genIntList}$_\texttt{safe}$, capturing the fact that
it always returns the empty list: %
\vspace{-.4cm}\par\nobreak %
{ \small
\begin{align*}
  \ot{n}{\Code{int}}{\Code{n} \geq 0} \rightarrow \ot{\Code{l}}{\Code{int~list}}{\Code{len(l)} = 0}
\end{align*}
}
\noindent Thus, any value of the type
$\ot{l}{\Code{int~list}}{\Code{len(l)} \neq 0}$ must fall outside the range
of \ocamlinline{genIntList}$_\texttt{safe}$.

To carry out our experiment, we applied a safety-focused repair
strategy that replaces \synthesize{} with the repair that an
off-the-shelf, safety-guided synthesizer~\cite{synquid} would generate
for each hole.  The hole corresponding to the base case of the sized
list benchmark yields the following synthesis goal, for example:
\vspace{-.4cm}\par\nobreak %
{ \small
\begin{align*}
 \Code{n}:\ot{n}{\Code{int}}{\Code{n} = 0} \vdash \hole{}_1 ~:~ \ot{\Code{l}}{\Code{int~list}}{\Code{len(l)} \leq \Code{n}}
\end{align*}
}
\noindent
To approximate the relative completeness of \name{} versus a
safety-focused repair approach, we first examined each safety-repaired
generator and came up with a standard refinement type for its
outputs. We then generated 20k values from the \name{}-repaired
generator and used the qualifier of this type to identify how many of
these outputs fall outside the range of the safety-repaired generator.
\autoref{fig:eval+safety} present the results of this experiment for
each of the list and tree variant from \autoref{tab:eval+RQ1}. The
taller the bar, the more (relatively) complete the \name{}-repaired
generator.

On these benchmarks, the completeness of the safety-focused repair
strategy tends to be an all-or-nothing proposition: in the case of the
sized list benchmark, for example, the safety-focused strategy's
prioritization of minimal terms yields generators that always return a
\ocamlinline{[]} term for six of the nine variants; this is precisely
the repair needed by three of these, however. In contrast, the
safety-focused strategy tends to be more effective when the set of
valid repairs are strongly constrained by the arguments of a
generator: the length of the output lists in the unique and duplicate
benchmarks is completely determined by its size parameter, for
example. The safe strategy is also effective when the required repair
is a constant: the required repair in the first two variants of depth
tree benchmark is a single \ocamlinline{Leaf} constructor, for
example.%

In contrast, the safe repair strategy tends to perform poorly when the
target precondition admits a number of safe repairs that are
relatively cheap: a cost function that employs Occam's razor, for
example, prioritizes small repairs that use variables or
constants. Even when the target property is quite restrictive, a
safety-focused strategy is biased towards repairs that produce values
of the right ``shape'', but whose contents do not vary much. When
repairing the first hole of \genEvensSk{}, for example, this strategy
is biased towards patches like \ocamlinline|[0]|; this term is
unlikely to explore code paths in the function under test that depend
on the contents of its list. Crucially, prioritizing completeness is
not simply a matter of equipping a safety-focused generator with a
different \emph{syntactic} cost function: the patch for the second
hole of \genEvensSk{} required synthesizing and joining two terms that
were both individually safe; the complete repair for the red-black
tree sketch similarly joins together multiple safe terms. Finding the
right combinations of terms to join requires \emph{semantic}
characterizations of both the missing coverage \emph{and} the coverage
provided by a candidate patch.

\paragraph{Discussion}
An important takeaway from both of these experiments is that while
completeness- and safety-focused repair strategies can yield useful
repairs in certain situations, their efficacy is highly dependent on
the particular problem, and there are many scenarios in which neither
approach performs well. When the target property is weak, a
completeness-focused repair can improve on the default generator,
while a safety-focused strategy tends to work well when the
specification is very stringent. Neither approach tends to do well when
the property falls somewhere between these two extremes, e.g., our BST
and red-black tree examples. Prioritizing the minimal
coverage-complete repair enables \name{} to produce repaired functions
that generate useful inputs without omitting any interesting values.

\subsection{Sensitivity to Set of Components (RQ4)}
\label{sec:RQ4}

As with any component-based synthesizer~\cite{ABJMRSSSTU+13,
  RRGLMRST+21, Trio}, the ability of \name{} to effectively find a
solution and the quality of that solution depends on the set of
components available to it.  This set of experiments investigates the
sensitivity of \name{} to the set of components it uses.

\paragraph{Extraneous Components} All of the benchmarks in
\autoref{sec:RQ1} used a common set of components based on the base
type of the target generator, so all of the list and tree benchmarks
included strictly more components than were needed to make the
coverage complete.  In order to measure how this inclusion of extra
components impacted \name{}'s ability to find a solution, this
experiment examines how much overhead such components added to the
synthesis time over a set that contains exactly the components needed
to precisely repair a sketch.

\begin{table}[t]
    \renewcommand{\arraystretch}{0.8}
    \caption{\small Results of adding superfluous components to
      synthesis benchmarks from \autoref{tab:eval+RQ1}. The columns
      show: \#Components (total, removed), \#Queries (total, \%
      change), \#Terms (total, \% change), and the synthesis time
      (total time, \% change).}
  \label{tab:excess_components}
  \centering \vspace*{-.05in} \footnotesize
\pgfplotstableread[col sep=comma]{fig/data/excess_comparison_combined.csv}\datatable
\centering \vspace*{-.05in} \footnotesize
\pgfplotstabletypeset[
 columns/Benchmark/.style={string type, column type=l},
 columns/Components/.style={string type, column type=l, column name={\#Components (removed)}},
 columns/Queries/.style={string type, column type=l, column name={\#Queries (\% change)}},
 columns/Terms/.style={string type, column type=l, column name={\#Terms (\% change)}},
 columns/Time/.style={string type, column type=l, column name={Time (sec, \% change)}},
 every head row/.style={before row=\toprule, after row=\midrule},
 every last row/.style={after row=\bottomrule},
 every nth row={4}{after row=\midrule},
]{\datatable}
\end{table}

\autoref{tab:excess_components} presents the results of these
experiments. For most of these experiments, the extra components
imposed relatively modest overhead to the performance of \name{} that
is roughly commensurate with the number of components
added.\footnote{The slight \emph{improvement} in the synthesis time of
  two of our smallest benchmarks can be attributed to the sensitivity
  of the underlying SMT solver to perturbations in input
  queries~\cite{SMT+Instability}.} There are a couple of reasons for
this: \name{} adopts the standard technique of using types to filter
out any ill-typed terms~\cite{FCD+15, myth, Trio} and prioritizes
lower-cost terms, limiting the number of additional terms the extra
components cause \name{} to enumerate before it finds a precise
solution. In addition, the bulk of the SMT queries in \synthesize{}
are limited to terms in the candidate pool, which does not include
many of the newly added terms. The one exception is our BST benchmark,
where the extraneous components lead to considerably more enumerated
terms and SMT queries. Here, the full set of components generates
several integer terms with similar costs and slightly different
coverages. Intuitively, because the solution uses one of these
components, \name{} has to perform more comparisons to identify which
of these variants to include in the final patch. In addition, the
recursive call of \ocamlinline|genBST| requires multiple arguments
with the same base types as the extraneous components, resulting in
several additional terms comprised of applications of the recursive
call to different arguments.

\begin{figure}[t!]
  \begin{subfigure}[t]{0.42\textwidth}
    \footnotesize
    \begin{ocaml}[fontsize=\footnotesize]
let rec even_list_gen s =
 if sizecheck s then
!\textcolor{red}{-}!  (int_gen() * 2) :: []
!\textcolor{green}{+}!  int_gen() :: []
 else if bool_gen() then
!\textcolor{red}{-}!  (int_gen() * 2) :: []
!\textcolor{green}{+}!  int_gen() :: []
 else
!\textcolor{red}{-}!  (int_gen() * 2) :: even_list_gen (s-1)
!\textcolor{green}{+}!  int_gen() :: even_list_gen (s-1)
    \end{ocaml}
    \caption{Repair for Even List sketch with no
      \ocamlinline|*|
      component}
    \label{fig:even+list+missing}
  \end{subfigure}\hfill
  \begin{subfigure}[t]{0.52\textwidth}
    \footnotesize
    \begin{ocaml}[fontsize=\footnotesize]
let rec rbtree_gen inv color h =
 if sizecheck h then
  if color then Rbtleaf
  else if bool_gen () then
!\textcolor{red}{-}!   Rbtleaf
!\textcolor{green}{+}!   rbtree_gen h true h
    else
      Rbtnode (true, Rbtleaf, int_gen (), Rbtleaf)
   (* ... rest omitted for space ... *)
 !\phantom{foo}!
    \end{ocaml}
    \caption{Repair for Red-Black Tree 3 with no
      \ocamlinline|Rbtleaf| component}
    \label{fig:rbtree+missing}
  \end{subfigure}
  \vspace{-.2cm}
  \caption{Diffs of a generator with a perfect repair and the variant
    \name{} finds when it lacks a component needed by that repair.}
  \label{fig:component+missing+examples}
  \vspace{-.2cm}
\end{figure}

\paragraph{Insufficient Components} Our second set of experiments
investigates the quality of the solution found by \name{} when it
cannot construct a precise repair, and effectively probe the ability
of \name{}'s ``best effort'' extraction mechanism to improve upon the
completeness-focused repair strategy from the previous experiment. For
this investigation, we examined each of the 44 benchmarks from
\autoref{tab:eval+RQ1} and then removed a key component used by the
original (complete) generator.

On almost half of the deliberately sabotaged versions of these
benchmarks (21/44) , \name{} was still able to produce a different
result than the completeness-focused repair. For many of these,
though, the practical difference with the default repair was not that
significant: without access to the \ocamlinline|-| component in the
duplicate list benchmark, for example, \name{} generates the repair %
\ocamlinline|x :: list_gen()| instead of %
\ocamlinline|x :: duplicate_list_gen (s-1) x|; the former is only a
marginal improvement over \ocamlinline|list_gen()|. On the other hand,
\name{} was able to produce more meaningfully useful repairs for the
two benchmarks shown in \autoref{fig:component+missing+examples}.  If
it is not supplied with the multiplication operator, \name{} builds
the repair shown in \autoref{fig:even+list+missing} for the sketch of
the even list generator --- while not perfect, this generator at least
produces lists with the required shape. \autoref{fig:rbtree+missing}
presents the repair found by \name{} for a red-black tree variant that
has a hole in one of its base cases when it is not provided the leaf
constructor. Here, \name{} somewhat cleverly makes a recursive call using
 \ocamlinline|h| --- which is guaranteed to be less than its recursive argument \ocamlinline|inv|
at this hole --- relying on the coverage provided by the other base cases
to repair the hole. As these two examples demonstrate, \name{} is
capable of finding non-trivial repairs even when it is not able to
construct a precise solution, but the utility of those repairs also
depends on the set of components available to it.

\paragraph{Discussion}
These experiments show that, as with other bottom-up synthesis
techniques, the set of components available to \name{} impacts both
its performance when constructing a repair and the quality of the
solution it finds--- hence the restricted set of components in the
STLC case study in \autoref{sec:RQ1}. On the one hand, the results of
the first experiment indicate that while \name{}'s current strategy
for prioritizing and filtering enumerated terms makes it reasonably
robust to extraneous components, it is still more effective when
equipped with a more targeted set of components. On the other hand,
the second experiment demonstrates that \name{}'s ability to construct
useful repairs when it cannot find a precise solution also depends on
the components it is provided. Taken together, these experiments
suggest that incorporating more advanced techniques from the program
synthesis community~\cite{Duet, Trio, FlashFill++, bustle, probe} for
curating the set of components used by \name{} is an important future
direction.

\subsection{Comparison with Dynamic Test Input Generation (RQ5)}
\label{sec:dyn+gen}
\begin{figure}[t]
  \centering
  \includegraphics[scale=.42]{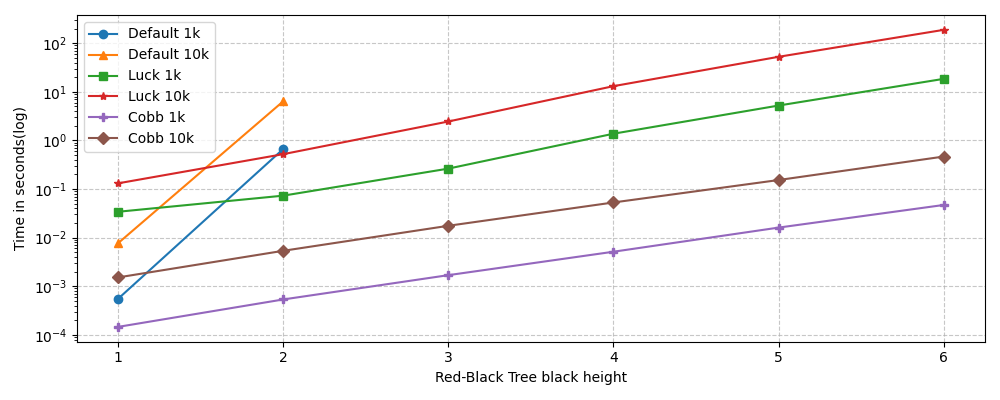}
  \vspace{-.25cm}
  \caption{Time needed to generate 1k and 10k valid red-black trees
    using \name{}-repaired generators, Luck~\cite{LGHHPX+17}, and a
    default `randomly generate and filter' approach.}
  \label{fig:luck+eval}
  \vspace{-.25cm}
\end{figure}

The ultimate goal of \name{} is to use symbolic reasoning to
statically ensure that the set of values enumerated by a generator
aligns with the complete set of values that meet the precondition of a
function under test. One alternative approach is to instead use a
theorem prover to dynamically generate these values during
testing~\cite{LGHHPX+17, Target}. To evaluate these alternative styles
of test generation, this experiment compares \name{} with
Luck~\cite{LGHHPX+17}, a tool which queries a constraint solver to
produce values that satisfy a user-defined predicate written in a
DSL. \autoref{fig:luck+eval} reports the time needed for Luck to
generate 1k and 10k red-black trees, respectively, against the time
needed for \name{} to generate the same number of trees. As a
baseline, the figure also reports the time needed by to produce 1k and
10k valid red-black trees by running the default generator in
`generate and filter' loop. We use an upper time limit of 5 minutes
for all the experiments; only the baseline approach exceeds this
bound.

While not a perfect apples-to-apples comparison--- among other
things,\footnote{Unlike Luck, \name{}-repaired generators are not
  guaranteed to produce unique values, although the latter's use of
  \Code{int\_gen()} to produce (signed 63-bit) integers means they are
  statistically unlikely to produce duplicate trees.} these generators are
implemented in different languages and frameworks--- these experiments
confirm the conclusions of \citet{LGHHPX+17} that run-time constraint
checking imposes (at least) an order of magnitude amount of overhead
over a generator that does not solve constraints at runtime. One
takeaway from this experiment is that the overhead of dynamic
constraint solving quickly matches the overhead required by our static
repair approach-- it takes \name{} roughly a minute to repair the
red-black tree sketch, which is roughly the amount of time needed to
generate 10k red-black trees with a black height of at least 5. Given
that a repaired generator can be run without any additional constraint
solving, the overhead required by the static repair approach seems
reasonable for settings in which generators are repeatedly used, e.g.,
when using PBT in a CI setting~\cite{GCDPH+24}.

\subsection{Discussion and Limitations}
\label{sec:limitations}

While our sized generators are complete for an arbitrary size bound,
the experiments in Sections \ref{sec:eval+repair} and
\ref{sec:dyn+gen} use a more limited range of bounds when
generating values, as is common in the literature. All of our list
benchmarks use QCheck's built-in generator for natural numbers,
\ocamlinline{nat_gen()}. This generator produces integers between 0
and 10000, and its distribution of outputs is skewed towards smaller
values. The bound in our tree benchmarks limits the height of the
tree, which bounds the number of nodes in a tree at
$O(2^{n+1}-1)$. Simply using \ocamlinline{nat_gen()} for these
benchmarks can generate very large trees, so these benchmarks instead
use a height between 0 and 12, chosen at random; this range is also
used in prior works~\cite{LGHHPX+17}.

As mentioned in \autoref{sec:implementation}, \name{} only guarantees
that a value can be generated with non-zero probability, and says
nothing more about the likelihood that it will be generated. As a
consequence, \name{} does not ensure that a repaired generator is
fair, i.e., that every value can be produced with uniform
probability. In our experiments, the distribution of a repaired
generator's outputs largely depends on the structure of the original
generator. Reasoning about the fairness of repaired generators and
repairing unfair generators is an interesting direction for future
work.

\section{Related Work}
\label{sec:related}

\paragraph{Generating Data Meeting Sparse Preconditions}
A number of works have considered how to effectively generate data
satisfying a sparse precondition. The proposed solutions can be
roughly categorized as either \emph{dynamic} and \emph{static}
approaches. \emph{Dynamic approaches} attempt to directly ensure the
validity of inputs as they are being generated~\cite{Korat, GGJKKM+10,
  CDP+14, Target, LGHHPX+17}, typically by relying on run-time
constraint solving. Like \name{}, Target~\cite{Target} uses refinement
types to define the space of valid inputs. To generate values,
however, Target queries an SMT solver for a model satisfying the type
qualifier, and then converts the model into a value in the target
language. To generate additional values, the SMT query is updated to
explicitly exclude any models that have already been found. Another
particularly popular strategy is to directly leverage the definition
of the target precondition, lazily concretizing the value being
generated in a way that ensures the constraint is satisfied and
backtracking when constraints become unsatisfiable~\cite{Korat,
  GGJKKM+10, CDP+14, LGHHPX+17}, similar to the idea of narrowing in
logic programming languages. The completeness of dynamic approaches is
typically tied to the completeness of the underlying constraint
solver: as long as the solver can return any satisfying value, so can
the input generator. The need to solve constraints at run time imposes
considerable overhead however; as discussed in \autoref{sec:dyn+gen},
dynamic approaches can be orders-of-magnitude slower than their static
counterparts, particularly when the target property is complex.

\name{} instead adopts a \emph{static approach}. Static generation
techniques avoid run-time constraint solving, and instead seek to
construct input generators that are sound and complete \emph{by
  construction}. Closely related to \name{} is a line of work that
automatically builds generators for the QuickChick PBT
framework~\cite{quickchick} by compiling inductively defined relations
into efficient generators~\cite{LPP18, PEL+22}. This pipeline uses a
translation validation approach~\cite{PSS+98} to ensure the
correctness of the resulting generators, producing formal proofs of
their soundness and completeness in the Coq/Rocq proof
assistant. Unlike \name{}, which is agnostic to how the target
property is defined, these works impose a strict requirement that the
target precondition be defined as an inductive proposition in
Coq/Rocq, although recent work has considered how this restriction can
be somewhat relaxed by, e.g., composing different inductive relations
into a single unified definition~\cite{merging}. This restriction is
used to produce generators that closely follow the definition of the
proposition itself; \name{}, in contrast, is able to synthesize and
repair arbitrary programs that supply the desired coverage.

\paragraph{Generating a Good Distribution of Data}
An orthogonal problem to coverage is the question of the
\emph{distribution} of outputs produced by a generator: a generator
for trees that produces \ocamlinline{Leaf} nodes 90\% of the time is
less useful than one whose outputs are uniformly distributed, for
example. A few tools have been proposed for statically reasoning about
the distribution of a generator's outputs: one such example is
Feat~\cite{DJW+12}, is a library for writing enumerators for datatypes
that are guaranteed to produce a uniform distribution of the values of
an algebraic datatype. The \textsc{Dragen} tool~\cite{MRH+18, MR+19},
in contrast, uses a mathematical model to statically estimate the
distribution of constructors produced by a QuickCheck generator built
from its \Code{frequency} combinator, and uses those estimates to
adjust the arguments of \Code{frequency} to achieve a more desirable
distribution. Extending \name{} to account for the distribution of a
generator is an interesting direction for future work.

\paragraph{Program Synthesis}
Automatically deriving programs from logical specifications of their
behavior has been a goal of the programming synthesis community for
almost half a century~\cite{MW+79}. The overwhelming majority of
program synthesis techniques use specifications that overapproximate
the set of desired behaviors~\cite{ABJMRSSSTU+13, fiat, suslik,
  cobalt}, including those that also use refinement types to specify
program behaviors~\cite{synquid, GJJZWJP+19, hoogleplus}. The
specifications used by \name{}, in contrast, stipulate an
underapproximation of the desired behaviors. This impacts \name{}'s
repair algorithm, which combines partial solutions which do not
individually satisfy the target specification, to build a complete
solution. The sets of input-output examples used by inductive
synthesis, or programming-by-example (PBE),
techniques~\cite{Recprogsyn, frangel, Stun, myth, FOWZ+16, Burst,
  Duet, YRS+23, Trio} also underapproximate the target program's
behavior, although they do so much less comprehensively than coverage
types. While similar to the bottom-up term enumeration strategy
employed in PBE systems, \name{}'s \synthesize{} procedure is able to
take advantage of the more complete approximation provided by coverage
types to, e.g., recursively call the generator being repaired before
its full definition is known~\cite{Burst, YRS+23}.

\paragraph{Automated Program Repair}
The goal of automated program repair (APR) is to automatically patch a
buggy program with minimal user effort~\cite{LPR+19}. Most APR
approaches use test suites to identify buggy behaviors: a valid patch
is one that causes a program that was failing some tests to instead
pass its suite. A notable exception is the work of \citet{LB+12},
which defines a ``good'' repair as one that decreases the number of
statically detected assertion failures in a program without
introducing any new ones. A major challenge in APR is finding repairs
that generalize beyond a particular failing test~\cite{SBLB+15}, a
problem that \name{} avoids thanks to the strong correctness
specifications provided by coverage types. Like \name{}, several APR
techniques rely on program synthesis to generate candidate
repairs. \citet{NQRC+13}, for example, employ symbolic execution to
identify path constraints that cause tests to succeed or fail. These
constraints are used as a safety specification for the target repair,
which is then generated using component-based synthesis. An
alternative strategy is to use angelic execution~\cite{MYR+16,
  LCLLV+17} to identify concrete values that can be used to help a
program pass a failing test; finding a patch that generates these
values is an instance of a PBE problem. As previously discussed, the
program synthesis techniques employed by both strategies use
specifications that are fundamentally different from \name{}'s.

\section{Conclusion}
\label{sec:conclusion}

When using a property-based testing framework to automatically test a
program that has a restrictive or sparse precondition, users are
typically forced to manually write a function that effectively
generates values of interest. Alongside the additional burden this
imposes on users of PBT frameworks, this process is also error-prone,
as generators can be both unsound, producing values that do not meet
the target precondition, and incomplete, incapable of producing all
the values that meet the precondition. This paper presents a technique
for detecting and repairing incomplete test input generators,
leveraging coverage types to characterize the set of missing test
values and the coverage provided by candidate repairs. Our repair
technique uses a novel coverage-type guided enumerative synthesis
algorithm to generate candidate repairs, employing a lattice structure
to store partial solutions so that they can be efficiently queried and
combined to build a complete repair. We have implemented a repair tool
for OCaml input generators, called \name{}, and have used it to repair
a diverse suite of benchmarks drawn from the PBT literature. Our
experiments demonstrate that \name{} can also be effective as a
sketch-based synthesis tool for test input generators, suggesting its
potential for further reducing a possible point of friction for users
of PBT frameworks.

\section*{Data-Availability Statement}
An artifact containing the source code for \name{}, its dependencies,
our suite of benchmark programs, and the scripts used to produce our
experimental results is publicly available on
Zenodo~\cite{cobb:artifact}.

\begin{acks}
  We thank the anonymous reviewers for their detailed comments and
  suggestions, Anxhelo Xhebraj for his help with the design and
  implementation of a preliminary version of Cobb, Francille Zhuang
  for her help with the evaluation of RQ3, and Kartik Sabharwal for
  his guidance in debugging SMT solver performance and behaviors.
  This research was partially supported by the National Science
  Foundation under Grant CCF-SHF 2321680.
\end{acks}

\bibliographystyle{ACM-Reference-Format}
\bibliography{paper}

\ifdefined\fullmode
\newcommand{\showappendix}{true}
\else
\fi

\appendix
\newpage
\section{Operational Semantics}
\label{sec:tech+semantics}
\begin{figure}[hb]
 {\small
  {\normalsize
\begin{flalign*}
 &\text{\textbf{Operational Semantics }} & \fbox{$e \hookrightarrow e$}
\end{flalign*}
}
\\ \
\begin{prooftree}
  \hypo{v \models \phi}
  \infer1[\textcolor{DeepGreen}{\textsc{EHole}}]
  {
    \hole{}~:~\nuut{b}{\phi} \hookrightarrow v}
\end{prooftree}
\quad
\begin{prooftree}
\hypo{op\ \overline{v} \equiv v_y }
\infer1[\textsc{EAppOp}]{
\zlet{y}{op\ \overline{v}}{e} \hookrightarrow e[y\mapsto v_y]
}
\end{prooftree}
\\ \ \\ \ \\
\begin{prooftree}
\hypo{e_1 \hookrightarrow e_1'}
\infer1[\textsc{ELetE1}]{
\zlet{y}{e_1}{e_2} \hookrightarrow \zlet{y}{e_1'}{e_2}
}
\end{prooftree}
\quad
\begin{prooftree}
\hypo{}
\infer1[\textsc{ELetE2}]{
\zlet{y}{v}{e} \hookrightarrow e[y\mapsto v]
}
\end{prooftree}
\\ \ \\ \ \\
\begin{prooftree}
\hypo{}
\infer1[\textsc{ELetAppLam}]{
\zlet{y}{\zlam{x}{t}{e_1}\ v_x}{e_2} \hookrightarrow \zlet{y}{e_1[x\mapsto v_x]}{e_2}
}
\end{prooftree}
\\ \ \\ \ \\
\begin{prooftree}
\hypo{}
\infer1[\textsc{ELetAppFix}]{
\zlet{y}{\zfix{f}{t}{x}{t_x}{e_1}\ v_x}{e_2} \hookrightarrow
\zlet{y}{(\zlam{f}{t}{e_1[x\mapsto v_x]}) \ (\zfix{f}{t}{x}{t_x}{e_1})}{e_2}
}
\end{prooftree}
\\ \ \\ \ \\
\begin{prooftree}
\hypo{}
\infer1[\textsc{EMatch}]{
\match{d_i \ \overline{v_j}} \overline{d_i\ \overline{y_j} \to e_i} \hookrightarrow e_i[\overline{y_j \mapsto v_j}]
}
\end{prooftree}
}
\caption{Small Step Operational Semantics of \langname{}.}
    \label{fig:semantics}
\end{figure}

\autoref{fig:semantics} gives the reduction rules for \langname{}'s
small standard operational semantics.

\newpage
\section{Type System}
\label{sec:tech+typing}

\begin{figure}[hb]
{\small
{\normalsize
\begin{flalign*}
 &\text{\textbf{Well-Formedness }} & %
 \fbox{$\Gamma \vdashunder \tau$}
\end{flalign*}
}
\\ \
\begin{prooftree}
\hypo{
\parbox{110mm}{\center
$\Gamma \equiv \overline{x_i{:}\nuot{b_{x_i}}{\phi_{x_i}}, y_j{:}\nuut{b_{y_j}}{\phi_{y_j}}, z{:}(a{:}\tau_a{\shortrightarrow}\tau_b)}$ \\
  $(\overline{\forall x_i{:}b_{x_i}, \exists y_j{:}b_{y_i}},\forall \nu{:}b, \phi) \text{ is a Boolean predicate}$ \quad
  $\forall j, \exn \not\in \denotation{\nuut{b_{y_j}}{\phi_{y_j}}}_{\Gamma}$
}
}
\infer1[\textsc{WfBase}]{
\Gamma \vdashunder \nuut{b}{\phi}
}
\end{prooftree}
\\ \ \\ \ \\
\begin{prooftree}
\hypo{\Gamma, x{:}\nuot{b}{\phi} \vdashunder \tau}
\infer1[\textsc{WfArg}]{
  \Gamma \vdashunder x{:}\nuot{b}{\phi}{\shortrightarrow}\tau
}
\end{prooftree}
\quad
\begin{prooftree}
\hypo{\Gamma \vdashunder (a{:}\tau_a{\shortrightarrow}\tau_b)}
\hypo{\Gamma \vdashunder \tau}
\infer2[\textsc{WfRes}]{
\Gamma \vdashunder (a{:}\tau_a{\shortrightarrow}\tau_b){\shortrightarrow}\tau
}
\end{prooftree}
\\
{\normalsize
\begin{flalign*}
 &\text{\textbf{Subtyping }} & \fbox{$\Gamma \covervdash \tau_1 <: \tau_2$}
\end{flalign*}
}
\\ \
\begin{prooftree}
\hypo{
\parbox{40mm}{\raggedright
$\denotation{\nuut{b}{\phi_1}}_{\Gamma} \subseteq \denotation{\nuut{b}{\phi_2} }_{\Gamma}$
}}
\infer1[\textsc{SubUBase}]{
\Gamma \covervdash \nuut{b}{\phi_1} <: \nuut{b}{\phi_2}
}
\end{prooftree}
\quad
\begin{prooftree}
\hypo{
\parbox{37mm}{\raggedright
$\denotation{\nuot{b}{\phi_1}}_{\Gamma} \subseteq \denotation{\nuot{b}{\phi_2}}_{\Gamma}$
}}
\infer1[\textsc{SubOBase}]{
\Gamma \covervdash \nuot{b}{\phi_1} <: \nuot{b}{\phi_2}
}
\end{prooftree}
 \\ \ \\ \ \\
\begin{prooftree}
\hypo{
\parbox{47mm}{\raggedright
$\Gamma \covervdash \tau_{21} <: \tau_{11}$ \quad
$\Gamma, x{:}\tau_{21} \covervdash \tau_{12} <: \tau_{22}$
}
}
\infer1[\textsc{SubArr}]{
\Gamma \covervdash x{:}\tau_{11}{\shortrightarrow}\tau_{12} <: x{:}\tau_{21}{\shortrightarrow}\tau_{22}
}
\end{prooftree}
{\normalsize
\begin{flalign*}
 &\text{\textbf{Disjunction }} & \fbox{$\Gamma \covervdash \tau_1 \lor \tau_2 = \tau_3$}
\end{flalign*}
}
\begin{prooftree}
\hypo{
\denotation{\tau_1}_{\Gamma} \cap \denotation{\tau_2}_{\Gamma} = \denotation{\tau_3}_{\Gamma}}
\infer1[\textsc{Disjunction}]{
\Gamma \covervdash \tau_1 \lor \tau_2 = \tau_3
}
\end{prooftree}
}
\caption{Auxillary typing relations}\label{fig:aux-rules}
\end{figure}

\begin{figure}[!ht]
  {\small
{\normalsize
\begin{flalign*}
 &\text{\textbf{Typing }} & \fbox{$\Gamma \covervdash e : \tau$}
\end{flalign*}
}
\\ \
\begin{prooftree}
  \hypo{\Gamma \vdashunder \nuut{b}{\phi}}
  \infer1[\textcolor{DeepGreen}{\textsc{THole}}]{ \Gamma \covervdash
    \hole{}~:~\nuut{b}{\phi} : \nuut{b}{\phi} }
\end{prooftree}
\quad
\begin{prooftree}
\hypo{\Gamma \vdashunder \nuut{b}{\bot}}
\infer1[\textsc{TErr}]{
\Gamma  \covervdash \exn : \nuut{b}{\bot}
}
\end{prooftree}
\\ \ \\ \ \\
\begin{prooftree}
\hypo{\Gamma \vdashunder \S{Ty}(c)}
\infer1[\textsc{TConst}]{
\Gamma \covervdash c : \S{Ty}(c)
}
\end{prooftree}
\quad
\begin{prooftree}
\hypo{\Gamma \vdashunder \S{Ty}(op)}
\infer1[\textsc{TOp}]{
\Gamma \covervdash op : \S{Ty}(op)
}
\end{prooftree}
\\ \ \\ \ \\
\begin{prooftree}
\hypo{\Gamma \vdashunder \nuut{b}{\nu = x}}
\infer1[\textsc{TVarBase}]{
\Gamma \covervdash x : \nuut{b}{\nu = x}
}
\end{prooftree}
\quad
\begin{prooftree}
\hypo{\Gamma(x) = (a{:}\tau_a{\shortrightarrow}\tau_b) \quad \Gamma \vdashunder a{:}\tau_a{\shortrightarrow}\tau_b}
\infer1[\textsc{TVarFun}]{
\Gamma \covervdash x : (a{:}\tau_a{\shortrightarrow}\tau_b)
}
\end{prooftree}
\\ \ \\ \ \\
\begin{prooftree}
\hypo{
\Gamma, x{:}\tau_x \vdash e : \tau \quad
\Gamma \vdashunder x{:}\tau_x{\shortrightarrow}\tau
}
\infer1[\textsc{TFun}]{
\Gamma \vdash \lambda x{:}\eraserf{\tau_x}.e : (x{:}\tau_x{\shortrightarrow}\tau)
}
\end{prooftree}
\\ \ \\ \ \\
  \begin{prooftree}
\hypo{
\Gamma,\; x{:}\nuot{b}{\phi},\; f{:}\nuotxarr{x}{b}{\nu \prec x~\land~ \phi}\tau ~~\covervdash~~ e : \tau \quad
\Gamma \vdashunder \nuotxarr{x}{b}{\phi}\tau
}
\infer1[\textsc{TFix}]{
  \Gamma \vdash \zfix{f}{\tau}{x}{b}{e} ~~:~~ \nuotxarr{x}{b}{\phi}\tau
}
\end{prooftree}
\\ \ \\ \ \\
\begin{prooftree}
\hypo{
\parbox{27mm}{\center
$\emptyset \covervdash \tau <: \tau'$\quad
$\emptyset \covervdash e : \tau$ \quad $\Gamma \vdashunder \tau'$
}}
\infer1[\textsc{TSub}]{
\Gamma \covervdash e : \tau' }
\end{prooftree}
\quad
\begin{prooftree}
\hypo{
\parbox{33mm}{\center
$\Gamma \covervdash \tau' <: \tau$ \quad
$\Gamma \covervdash \tau <: \tau'$ \\
$\Gamma \covervdash e : \tau$ \quad
$\Gamma \vdashunder \tau'$
}}
\infer1[\textsc{TEq}]{
\Gamma \covervdash e : \tau' }
\end{prooftree}
\\ \ \\ \ \\
\begin{prooftree}
\hypo{
\parbox{40mm}{\center
$\Gamma \covervdash e : \tau_1$\quad
$\Gamma \covervdash e : \tau_2$\quad
$\Gamma \covervdash \tau_1 \lor \tau_2 = \tau$ \quad
$\Gamma \vdashunder \tau$
}}
\infer1[\textsc{TMerge}]{
\Gamma \covervdash e : \tau }
\end{prooftree}
\quad
\begin{prooftree}
\hypo{
\parbox{43mm}{\center
$\Gamma \covervdash e_x : \tau_x$\quad
$\Gamma, x{:}\tau_x \covervdash e : \tau$\quad
$\Gamma \vdashunder \tau$
}
}
\infer1[\textsc{TLetE}]{
\Gamma \covervdash \zlet{x}{e_x}{e} : \tau
}
\end{prooftree}
\\ \ \\ \ \\
\begin{prooftree}
\hypo{
\parbox{43mm}{\center
$\Gamma \covervdash op : \overline{\nuotxarr{a_i}{b_i}{\phi_i}}\tau_x$\quad
$\forall i, \Gamma \covervdash v_i : \nuut{b_i}{[\phi_i}$\quad
$\Gamma, x{:}\tau_x[\overline{a_i \mapsto v_i}] \covervdash e : \tau$ \quad
$\Gamma \vdashunder \tau$
}
}
\infer1[\textsc{TAppOp}]{
\Gamma \covervdash \zlet{x}{op\ \overline{v_i}}{e} : \tau
}
\end{prooftree}
\quad
\begin{prooftree}
\hypo{
\parbox{43mm}{\center
$\Gamma \covervdash v_1 : (\tau_1{\shortrightarrow}\tau_2){\shortrightarrow}\tau_x$\quad
$\Gamma \covervdash v_2 : \tau_1{\shortrightarrow}\tau_2$\quad
$\Gamma, x{:}\tau_x \covervdash e : \tau$ \quad
$\Gamma \vdashunder \tau$
}
}
\infer1[\textsc{TAppFun}]{
\Gamma \covervdash \zlet{x}{v_1\ v_2}{e} : \tau
}
\end{prooftree}
\\ \ \\ \ \\
\begin{prooftree}
\hypo{
\parbox{43mm}{\center
$\Gamma \covervdash v_1 : \nuotxarr{a}{b}{\phi}\tau_x$\quad
$\Gamma \covervdash v_2 : \nuut{b}{\phi}$\quad
$\Gamma, x{:}\tau_x[a \mapsto v_2] \covervdash e : \tau$ \quad
$\Gamma \vdashunder \tau$
}
}
\infer1[\textsc{TApp}]{
\Gamma \covervdash \zlet{x}{v_1\ v_2}{e} : \tau
}
\end{prooftree}
\quad
\begin{prooftree}
\hypo{
\parbox{55mm}{\center
$\Gamma \covervdash v : \tau_v$\quad
$\Gamma \vdashunder \tau$\quad
$\Gamma, \overline{{y}{:}\tau_y} \covervdash d_i(\overline{{y}}) : \tau_v$\quad
$\Gamma, \overline{{y}{:}\tau_y} \covervdash e_i : \tau$
}}
\infer1[\textsc{TMatch}]{
\Gamma \covervdash (\match{v} \overline{d_i\ \overline{y} \to e_i}) : \tau}
\end{prooftree}
}
    \caption{Full Typing Rules}\label{fig:full-type-rules}
\end{figure}

The full set of typing rules for \langname{} is shown in
\autoref{fig:full-type-rules}.

\subsection{Type Denotations}

Assuming a standard typing judgement for
basic types, $\emptyset \basicvdash e : t$,
a type denotation for a type $\tau$, $\denot{\tau}$, is a set of
closed expressions:
\begin{align*}
    &\llbracket \rawnuot{b}{\phi} \rrbracket &&\equiv \{ v ~|~ \emptyset \basicvdash v : b \land \phi[\nu\mapsto v] \}\\
    &\llbracket \urt{b}{\phi} \rrbracket &&\equiv \{ e ~|~ \emptyset \basicvdash e : b \land \forall v{:}b, \phi[\nu\mapsto v] \impl e \hookrightarrow^* v \}\\
    &\llbracket x{:}\tau_x\sarr\tau \rrbracket &&\equiv \{ f~|~ \emptyset \basicvdash f : \eraserf{\tau_x\sarr\tau} \land
    \forall v_x \in \llbracket \tau_x \rrbracket \impl f\ v_x \in  \llbracket \tau[x\mapsto v_x ] \rrbracket \}
\end{align*}
In the case of an overapproximate refinement type, $\rawnuot{b}{\phi}$,
the denotation is simply the set of all values of type $b$
whose elements satisfy the type's refinement predicate ($\phi$), when
substituted for all free occurrences of $\nu$ in $\phi$.\footnote{The
  denotation of an overapproximate refinement type is more generally
  $\{ e{:}b ~|~ \emptyset \vdash e: b \land \forall
  v{:}b, e \hookrightarrow^* v \impl \phi[x\mapsto v] \}$.
  However, because such types are only used for function parameters,
  and our language syntax only admits values as arguments, our
  denotation uses the simpler form.}  Dually, the denotation of an
underapproximate coverage type is the set of expressions that evaluate
to $v$ whenever $\phi[\nu \mapsto v]$ holds, where $\phi$ is the
type's refinement predicate.  Thus, every expression in such a
denotation serves as a witness to a feasible, type-correct,
execution.  The denotation for a function type is defined in terms of
the denotations of the function's argument and result in the usual
way, ensuring that our type denotation is a logical predicate.

The denotation of a
refinement type $\tau$ under a type context $\Gamma$ (written
$\denot{\tau}_{\Gamma}$) is:\footnote{In the last case, since $\hat{e}_x$ may
  nondetermistically reduce to multiple values, we employ
  intersection (not union), similar to the \textsc{Disjunction}
  rule.}$^,$
{\small
\begin{alignat*}{2}
    \denot{\tau}_{\emptyset} &\equiv \denot{\tau}
    \\ \denot{\tau}_{x{:}\tau_x, \Gamma} &\equiv \{ e ~|~
    \forall v_x \in \denot{\tau_x}. \zlet{x}{v_x}{e} \in
    \denot{\tau [x\mapsto v_x]}_{\Gamma [x\mapsto v_x]} \} \hspace{1.8cm} \text{ if $\tau
      \equiv \rawnuot{b}{\phi}$} \\
      \denot{\tau}_{x{:}\tau_x,
      \Gamma} &\equiv \{ e ~|~ \exists \hat{e}_x \in
    \denot{\tau_x}. \forall e_x \in \denot{\tau_x}.
    \\ &\hspace{1.8cm} \zlet{x}{e_x}{e} \in \bigcap_{\hat{e}_x \hookrightarrow^*
      v_x} \denot{\tau
    [x\mapsto v_x]}_{\Gamma [x\mapsto v_x]} \}
    \hspace{1.7cm} \text{ otherwise}
\end{alignat*}}\noindent
The denotation of an overapproximate refinement type under a type
context is mostly unsurprising, other than our presentation choice to
use a let-binding, rather than substitution, to construct the expressions
included in the denotations.

We define the subset relation between the denotation of two refinement
types $\tau_1$ and $\tau_2$ under a type context $\Gamma$ (written
$\denotation{\tau_1}_{\Gamma} \subseteq \denotation{\tau_1}_{\Gamma}$)
as:
\begin{alignat*}{2}
    \denotation{\tau_1}_{\emptyset} \subseteq \denotation{\tau_2}_{\emptyset}&\equiv \denotation{\tau_1} \subseteq  \denotation{\tau_2}
    &\\ \denotation{\tau_1}_{x{:}\tau_x, \Gamma} \subseteq \denotation{\tau_1}_{x{:}\tau_x, \Gamma} &\equiv
    \forall v_x \in \denotation{\tau_x},\\
    &\qquad \denotation{\tau_1 [x\mapsto v_x]}_{\Gamma [x\mapsto v_x]} \subseteq
    \denotation{\tau_2 [x\mapsto v_x]}_{\Gamma [x\mapsto v_x]}
    &\quad \text{ if $\tau
      \equiv \nuot{b}{\phi}$} \\
       \denotation{\tau_1}_{x{:}\tau_x \Gamma} \subseteq \denotation{\tau_2}_{x{:}\tau_x \Gamma}
      &\equiv \exists \hat{e}_x \in
    \denotation{\tau_x}, \forall v_x, \hat{e}_x \hookrightarrow^*
      v_x \implies  \\
    &\qquad \denotation{\tau_1
    [x\mapsto v_x]}_{\Gamma [x\mapsto v_x]} \subseteq \denotation{\tau_2
    [x\mapsto v_x]}_{\Gamma [x\mapsto v_x]}
    &\text{ otherwise}
\end{alignat*}

The way we interpret the type context $\Gamma$ here is the same as the
definition of the type denotation under the type context, but we keep
the denotation of $\tau_1$ and $\tau_2$ as the subset relation under
the same interpretation of $\Gamma$, that is under the \emph{same}
substitution $[x\mapsto v_x]$. This constraint is also required by
other refinement type systems, which define the denotation of the type
context $\Gamma$ as a set of substitutions, with the subset relation
of the denotation of two types holding under the \emph{same}
substitution. However, our type context is more complicated, since it
has both under- and overapproximate types. The denotations of these
types use both existential and universal quantifiers, and cannot
simply be interpreted as a set of substitutions. Thus, we define a
subset relation over denotations under a type context to ensure the
same substitution is applied to both types.

\begin{theorem}[Type Soundness~\cite{poirot}]
  \label{cor:typed+complete}
  A well-typed test generator of type
  $ \covervdash f ~:~ \overline{x_i: \nuot{b_i}{\phi_i}}
  \shortrightarrow \nuut{b}{\phi} $, when applied to well-typed
  arguments $\overline{\covervdash v_i~:~\nuot{b_i}{\phi_i}}$, can
  evaluate every value satisfying $\phi[\overline{x_i\mapsto v_i}] $:
  $\forall v.\; \phi[\overline{x_i\mapsto v_i}, \nu\mapsto v] \impl
  f\; \overline{v_i} \hookrightarrow^* v$
\end{theorem}
\begin{proof}
We proceed by specializing the general type soundness theorem for coverage types~\cite{poirot} to our setting. The general theorem states that for any type context $\Gamma$, term $e$, and coverage type $\tau$, if $\Gamma \vdash e : \tau$, then $e \in \denot{\tau}_{\Gamma}$.

In our case, the generator $f$ is well-typed in the empty context:
\[
\covervdash f : \overline{x_i: \nuot{b_i}{\phi_i}} \shortrightarrow \nuut{b}{\phi}
\]
and each argument $v_i$ is well-typed:
\[
\forall i.\; \covervdash v_i : \nuot{b_i}{\phi_i}
\]
By the soundness theorem, this means:
\begin{align*}
  &f \in \denotation{\overline{x_i: \nuot{b_i}{\phi_i}} \shortrightarrow \nuut{b}{\phi}} \\
  &\forall i.\; v_i \in \denotation{\nuot{b_i}{\phi_i}}
\end{align*}

By the definition of the denotation of function types, we have:
\[
\forall \overline{v_i} \in \denotation{\nuot{b_i}{\phi_i}}. \quad f~\overline{v_i} \in \denotation{\nuut{b}{\phi}\overline{[x_i \mapsto v_i]}}
\]
That is, for any choice of well-typed arguments $\overline{v_i}$, the application $f~\overline{v_i}$ is in the denotation of the result type under the substitution $[x_i \mapsto v_i]$.

Now, by the definition of the denotation of the coverage base type $\nuut{b}{\phi}$, we have:
\[
\denotation{\nuut{b}{\phi}\overline{[x_i \mapsto v_i]}} = \{ e \mid \forall v.\; \phi[\overline{x_i \mapsto v_i}, \nu \mapsto v] \implies e \hookrightarrow^* v \}
\]
Therefore, for any $v$ such that $\phi[\overline{x_i \mapsto v_i}, \nu \mapsto v]$ holds, it must be that $f~\overline{v_i} \hookrightarrow^* v$.

This is precisely the statement of the theorem:
\[
\forall v.\; \phi[\overline{x_i \mapsto v_i}, \nu \mapsto v] \implies f~\overline{v_i} \hookrightarrow^* v
\]
Thus, the result follows directly.
\end{proof}

\newpage
\section{Abduction Algorithm}
\label{sec:tech+abduction}

\begin{algorithm}[ht]
  \small
  \Params{$\Gamma$: typing context, %
    $\nuut{b}{\psi_\texttt{cur}}$: current coverage, %
    $\nuut{b}{\psi}$: target coverage} %
  \Output{Formula $\psi_\texttt{need}$ such that $\Gamma \covervdash \nuut{b}{\psi_\texttt{cur} \lor \psi_\texttt{need}} <: \nuut{b}{\psi}$}
  \lIf{$\Gamma \covervdash \nuut{b}{\phi_\texttt{cur}} <: \nuut{b}{\phi}$}{
    \Return{$\bot$}
  }
  $\Psi \leftarrow \bigcup\limits_{\phi \subseteq \Phi}\{\psi ~|~ \psi =
  \bigwedge\limits_{\alpha \in \phi} \alpha \land \bigwedge\limits_{\alpha \in \Phi -
    \phi} \lnot \alpha \} $\; %

  $\Psi^- \leftarrow \{\psi \in \Psi ~|~ \psi_\texttt{cur} \implies \psi \}$;
  $\Psi^? \leftarrow \Psi - \Psi^-$;
  $\Psi^+ \leftarrow \emptyset$;
  $\psi_\texttt{need} \leftarrow \top$\;
  \While{$\exists\, \psi_? \in \Psi^?$}{
    $\Psi^? \leftarrow \Psi^? \setminus \{\psi_?\}$\;
    $\psi_\texttt{need} \leftarrow \learn(\Psi^+ \cup \Psi^?, \Psi^- \cup \{\psi_?\})$\;
    \uIf{$\Gamma \covervdash \nuut{b}{ \psi_\texttt{cur} \lor \psi_\texttt{need}} <: \nuut{b}{\psi}$}{
      $\Psi^- \leftarrow \Psi^- \cup \{\psi_? \}$\;
    }
    \Else{
      $\Psi^+ \leftarrow \Psi^+ \cup \{\psi_?\}$\;
      $\psi_\texttt{need} \leftarrow \learn(\Psi^+, \Psi^- \cup \Psi^?)$\;
        \uIf{$\Gamma \covervdash \nuut{b}{ \psi_\texttt{cur} \lor \psi_\texttt{need}} <: \nuut{b}{\psi}$}{
        \Return{$\psi_\texttt{need}$}\;
        }
    }
  }
  \Return{$\psi_\texttt{need}$}\;
  \caption{Inferring missing coverage (\generalize{})}
  \label{algo:generalize}
\end{algorithm}

\autoref{algo:generalize} lists the \generalize{} subroutine that
\complete{} uses to infer $\psi_\texttt{need}$, the qualifier of the
type we use to capture an input generator's missing coverage. The
algorithm is parameterized over a set of atomic formulas, $\Phi$, and
take three arguments: the first argument, $\Gamma$, is the typing
context of the body of the generator, and the final two arguments
include the current and target coverage of the test input generator,
$\psi_\texttt{cur}$ and $\psi$ respectively. Given these inputs, the
goal of \generalize{} is to output the weakest formula
$\psi_\texttt{need}$ in the hypothesis space that ensures
$\Gamma \covervdash \nuut{b}{\phi_\texttt{cur} \lor
  \phi_\texttt{need}} <: \nuut{b}{\phi}$.

\generalize{} does so by adapting an existing algorithm for inferring
a maximally weak specification in the context of safety
verification~\cite{ZDDJ+21} to our coverage type setting.

This algorithm maintains three disjoint sets, $\Psi^-$, $\Psi^+$, and
$\Psi^?$, which cumulatively contain all conjunctions of the method
predicates in $P$ and their negations (line $2$). Intuitively, each
element of these sets is a formula that defines a distinct subset of
the generator's outputs: the formula
$\lnot \Code{empty}(l) \land \Code{hd}(l) = \Code{1}$, for example,
characterizes all nonempty lists that begin with \Code{1}.
The set $\Psi^+$ captures values that the input generator does not
output but needs to, $\Psi^-$ includes values that can safely omitted
from the output of the repaired generator, and $\Psi^?$ includes those
values which have not been definitively placed into either category.
\generalize{} initializes these sets by moving all values currently
covered by $\psi_\texttt{cur}$ into $\Psi^-$ and placing the remaining
elements into $\Psi^?$ (line 3). The candidate solution maintained by
\generalize{}, $\psi_\texttt{need}$, contains all the elements of
$\Psi^+$ and $\Psi^?$. The key invariant of \generalize{} is that the
disjunction of $\psi_\texttt{need}$ and $\psi_\texttt{cur}$ is always
a subtype of $\psi$, i.e., $\psi_\texttt{need}$ captures a superset of
the outputs that need to be added to a generator.

The algorithm's main loop (lines 4-13) attempts to place all the
members of $\Psi^?$ into either $\Psi^-$ or $\Psi^+$. Each iteration
of the loop checks if it is safe to move an element of $\Psi^?$,
$\psi_?$, to $\Psi^-$, adding it to $\Psi^+$ if not.  The loop first
uses an auxiliary function, \learn{},
to construct a candidate solution, $\psi_\texttt{need}$, that
distinguishes $\Psi^+$ and $\Psi^?$ from those of
$\Psi^- \cup \{\psi_?\}$ (line $5$). The loop then uses a subtype
check to see whether $\psi_\texttt{need}$ is still sufficient to
complete $\phi_\texttt{cur}$ (line 7), updating $\Psi^-$ if so (line
9). If not, $\psi_?$ is added to $\Psi^+$ and we check to see if all
the remaining elements of $\Psi^?$ can be safely moved to $\Psi^-$,
terminating if so (lines 11-13). If not, the loop continues until
$\Psi^?$ has no more elements.

\begin{theorem}[\generalize{} is Sound]
  \label{theorem:generalization+sound}
  Given a typing context $\Gamma$, %
  the current coverage $\nuut{b}{\psi_\texttt{cur}}$, %
  and the target coverage $\nuut{b}{\psi}$,
  \generalize($\Phi, \Gamma, \nuut{b}{\psi_\texttt{cur}},
  \nuut{b}{\psi}$) produces a
  $\psi_\texttt{need} \in \{ \psi' ~|~ \psi' =
  \bigvee(\bigwedge\overline{\phi} \land \bigwedge \overline{\lnot
    \phi}) \land \psi\}$ such that
  $\Gamma \covervdash \nuut{b}{\psi_\texttt{cur} \lor
    \psi_\texttt{need}} <: \nuut{b}{\psi}$.  Moreover,
  $\psi_\texttt{need}$ is a minimal solution in the solution space
  considered by \generalize{}:
  $\lnot \exists\, \psi' \in \{ \psi' ~|~ \psi' =
  \bigvee(\bigwedge\overline{\phi} \land \bigwedge \overline{\lnot
    \phi}) \land \psi\}.\; \Gamma \covervdash
  \nuut{b}{\psi_\texttt{cur} \lor \psi'} <: \nuut{b}{\psi} \land \psi'
  \implies \psi_\texttt{need}$.
\end{theorem}
\begin{proof}
  We first prove that the produced $\psi_\texttt{need}$ always satisfies $\Gamma \covervdash \nuut{b}{\psi_\texttt{cur} \lor \psi_\texttt{need}} <: \nuut{b}{\psi}$. The algorithm can only return in two cases: either at line 13, which guarantees that the subtyping relation holds, or at line 14, when all Boolean combinations of atomic formulas (i.e., $\bigwedge\limits_{\alpha \in \phi} \alpha \land \bigwedge\limits_{\alpha \in \Phi -
  \phi} \lnot \alpha$) are labeled as positive (i.e., $\Psi^+$) or negative (i.e., $\Psi^-$).
  If $\psi_\texttt{need}$ does not make the subtyping relation hold, then there exists a Boolean combination $\psi_\texttt{ex}$ such that
  \begin{align*}
    \Gamma \vdash \psi_\texttt{ex} \implies \psi \text{ and } \Gamma \not\vdash \psi_\texttt{ex} \implies \psi_\texttt{need} \lor \psi_\texttt{cur}
  \end{align*}\noindent
  Moreover, for any formula in the solution space, we have
  \begin{align*}
    \Gamma \not\vdash \psi_\texttt{ex} \implies \psi' \text{ implies that } \Gamma \not\covervdash \nuut{b}{\psi'} <: \nuut{b}{\psi}
  \end{align*}\noindent
  According to the definition of the learning procedure $\learn$, the formula $\psi_\texttt{ex}$ can only belong to the set $\Psi^-$. This can only happen on line 8, where $\psi_\texttt{need}$ makes the subtyping relation hold even when labeling $\psi_\texttt{ex}$ as negative, which conflicts with the assumption. Thus, the returned $\psi_\texttt{need}$ always satisfies $\Gamma \covervdash \nuut{b}{\psi_\texttt{cur} \lor \psi_\texttt{need}} <: \nuut{b}{\psi}$.

  Next, we prove minimality. If $\psi_\texttt{need}$ is not minimal, then there exists a formula $\psi'$ different from $\psi_\texttt{need}$ in the solution space such that
  \begin{align*}
    \Gamma \covervdash \nuut{b}{\psi_\texttt{cur} \lor \psi'} <: \nuut{b}{\psi} \land \psi' \implies \psi_\texttt{need}
  \end{align*}\noindent
  Then, there exists a Boolean combination $\psi_\texttt{max}$ such that
  \begin{align*}
    \Gamma \covervdash \psi_\texttt{max} \implies \psi_\texttt{need} \text{ and } \Gamma \not\covervdash \psi_\texttt{max} \implies \psi'
  \end{align*}\noindent
  According to the definition of the learning procedure $\learn$, the formula $\psi_\texttt{max}$ can only belong to the set $\Psi^+$. This can only happen on line 10, where $\psi_\texttt{need}$ fails to make the subtyping relation hold even when labeling $\psi_\texttt{max}$ as negative. This means that
  \begin{align*}
    \Gamma \covervdash \psi_\texttt{max} \implies \psi
  \end{align*}\noindent
  which is a contradiction with the assumption that $\Gamma \not\covervdash \psi_\texttt{max} \implies \psi'$ and $\Gamma \covervdash \nuut{b}{\psi_\texttt{cur} \lor \psi'} <: \nuut{b}{\psi}$. Thus, $\psi_\texttt{need}$ is a minimal solution.
\end{proof}

\newpage
\section{Localization Algorithm}
\label{sec:tech+localization}
\begin{algorithm}[ht]
  \small \Params{$\; s$: incomplete generator, %
    $\Gamma$: typing context, %
    $\nuut{b}{\phi_\texttt{need}}$: missing coverage } %
  \Output{$\;$ An updated sketch $i'$ containing $j$ holes and %
    a set of typing contexts for each hole
    $\overline{\Gamma_j \vdash \hole_j~:~\nuut{b}{\phi_j}}$} %
  \Match{$s$}{
    \lfCase{v}{\Return{$(v \oplus \hole{}~:~\nuut{b}{\phi_\texttt{need}}, \{\Gamma\vdash
        \hole{}~:~\nuut{b}{\phi_\texttt{need}}\})$}} %
    \lCase{\S{err}}{\Return{$(\hole{}~:~\nuut{b}{\phi_\texttt{need}}, \{\Gamma \vdash \hole{}~:~\nuut{b}{\phi_\texttt{need}}\})$}} %
    \lCase{$\hole{}~:~\nuut{b}{\phi}$}{\Return{($\hole{}, \{\Gamma
        \vdash \hole{}~:~\nuut{b}{\phi}, \Gamma \vdash \hole{}~:~\nuut{b}{\phi_\texttt{need}}\})$}} %
    \Case{$\zlet{x}{e_1}{e_2}$}{
      $(i_2, \Gamma') \leftarrow \localize{}(\Gamma; x:\typeInfer{}(
      \Gamma, e_1), e_2, \nuut{b}{\phi_\texttt{need}})$\;
      \Return{$(\zlet{x}{c_1}{i_2}, \Gamma')$}
    }
    \Case{$\zlet{x}{op\ \overline{v}}{e_2}$}{
      $(i_2, \Gamma') \leftarrow \localize{}(\Gamma; x:\typeInfer{}(
      \Gamma, op\ \overline{v}), e_2, \nuut{b}{\phi_\texttt{need}})$\;
      \Return{$(\zlet{x}{op\ \overline{v}}{i_2}, \Gamma')$} }
    \Case{$\zlet{x}{v\ v}{e_2}$}{
      $(i_2, \Gamma') \leftarrow \localize{}(\Gamma;
      x:\typeInfer{}(\Gamma, v\ v), e_2, \nuut{b}{\phi_\texttt{need}})$\;
      \Return{$(\zlet{x}{v\ v}{i_2}, \Gamma')$} }
    \Case{$\match{v} \overline{d_k\ \overline{y_k} \to e_k}$}{
      $\Gamma' \leftarrow \emptyset$\;
      \For{m $\in \mathsf{\{0, \ldots, k\}}$}{
        $\overline{{y}{:}\nuot{b}{\phi}}{\shortrightarrow}\nuut{b_m}{\psi_m}
        \leftarrow \S{Ty}(d_m)$\;
        $\Gamma_j' \leftarrow \overline{{y}{:}\nuot{b}{\phi}},\; a{:}\nuut{b_m}{\nu = v \land \psi_m}$\;
        $(i_m, \Gamma_m) \leftarrow
          \localize{}(\Gamma;\Gamma_j', e_j, \nuut{b}{\phi_\texttt{need}})$\;
        $\Gamma' \leftarrow \Gamma' \cup \Gamma_m$\;
      }
      \Return{$(\match{v} \overline{d_k\ \overline{y_k} \to i_k},\ \Gamma')$} }
  }
  \caption{Insert repair locations (\localize{})
  }
  \label{algo:localize}
\end{algorithm}

Starting from an initial typing context, $\Gamma$, \localize{}
recurses over the AST of the input program (line $1$). In the base
cases, a hole is inserted and the current typing context is attached
to it (lines 2-5). \localize{} replaces \Code{err} expressions with a
new hole, since errors never contribute any coverage (line 3). If
there is already a hole, \localize{} adds an additional hole with the
target coverage, and adds both holes to the set of repair locations.
In the recursive cases, $\Gamma$ is updated according to the typing
context used for each subterm in the corresponding typing rule: the
recursive call on line $10$, for example, extends the input typing
context with a binding for the result of $op\ \overline{v}$. When
applied to a $\mathtt{match}$ expression, \autoref{algo:localize}
recursively inserts holes into each branch (line $18$).

\newpage
\section{Correctness Proofs for \complete{}}
\label{sec:tech+repair}

Here are all the parameters that define the
  set of patches explored by \complete{}:
  \begin{itemize}
  \item A finite set of atomic formulas $\phi$ used by \generalize{};
  \item A collection of typed \textit{seeds} and \textit{components}
    that \synthesize{} uses to enumerate terms;
  \item A set of method predicates used in the types of those
    components and by \generalize{} to characterize missing coverage;
  \item Axioms characterizing the semantics of method predicates;
  \item A cost function used by \genExp{} to prioritize certain terms.
  \end{itemize}

The soundness proofs for \complete{} are based on the following three lemmas.

\begin{lemma}[Soundness of $\typeInfer{}$]
  For a given type context $\Gamma$ and incomplete program $s$, $\typeInfer{}(\Gamma, s)$ returns a coverage type $\nuut{b}{\psi}$ such that $\Gamma \covervdash s \; : \; \nuut{b}{\psi}$.
  \label{lemma:typeinfer-sound}
  \end{lemma}
\begin{proof}
  This follows directly from the type soundness theorem for coverage
  types~\cite{poirot}.
\end{proof}

\begin{lemma}[Soundness of $\localize{}$]
  For a given type context $\Gamma$, incomplete program $s$ that
  covers $\nuut{b}{\psi_\texttt{cur}}$, and coverage type
  $\nuut{b}{\psi_\texttt{need}}$, $\localize{}$ returns a new program
  sketch $s'$ and set of typed holes
  $\overline{\Gamma_j \vdash \hole{}_j : \nuut{b}{\psi_j}}$ such that
  for all terms $\overline{e_j}$ where
  $\overline{\Gamma_j \vdash e_j : \nuut{b}{\psi_j}}$,
  $s'[\overline{e_j}]$ is a repair of $s$ that covers
  $\nuut{b}{\psi_\texttt{cur} \lor \psi_\texttt{need}}$, i.e.,
  $\Gamma \vdash s'[\overline{e_j}]~:~\nuut{b}{\psi_\texttt{cur} \lor
    \psi_\texttt{need}}$ .
    \label{lemma:localize-sound}
    \end{lemma}
    \begin{proof}
    By induction on the structure of the program $s$.
\end{proof}

\begin{lemma}[Soundness of $\synthesize{}$]
  For a given type context $\Gamma$ and program sketch $s'$ with typed
  holes $\overline{\Gamma_j \vdash \hole{}_j : \nuut{b}{\psi_j}}$,
  such that for all terms $\overline{e_j}$ where
  $\overline{\Gamma_j \vdash e_j : \nuut{b}{\psi_j}}$,
  $\Gamma \vdash s'[\overline{e_j}] ~:~ \nuut{b}{\psi}$,
  $\synthesize{}$ returns a complete program with type
  $\nuut{b}{\psi}$: \\ %
  $\Gamma \vdash \synthesize(\Gamma, s', %
  \overline{\Gamma_j\vdash\hole{}_j~:~\nuut{b}{\psi_j}}, %
  \nuut{b}{\psi})~:~\nuut{b}{\psi}$.
  \label{lemma:synthesize-sound}
  \end{lemma}
  \begin{proof}
    \synthesize{} only exits immediately after the main loop if the
    current completion of $s'$ has the required coverage type:
    $\Gamma \vdash s'[\overline{e_j}] ~:~ \nuut{b}{\psi}$. Otherwise,
    for the i$^{th}$ hole in $s'$, the main body of $\synthesize{}$
    will have only produced a patch $e_i$ whose type precisely matches
    $\Gamma_i \vdash e_i : \nuut{b}{\psi_i}$ (lines 12-13), and the
    second loop will have filled any hole $\hole{}_k$ that remains
    with a term $e_k$ that is a subtype of $\nuut{b}{\psi_k}$, and
    thus $\Gamma_k \vdash e_k : \nuut{b}{\psi_k}$ -- in the worst
    case, this term will be the default generator for the base type
    \ocamlinline|b|, which is always included in our set of
    components.  Thus, if \synthesize{} exits on line 21,
    $\overline{\Gamma_j \vdash e_j : \nuut{b}{\psi_j}}$; by assumption
    \synthesize{} will return a completed sketch with the required coverage type
    $\Gamma \vdash s'[\overline{e_j}] ~:~ \nuut{b}{\psi}$.
\end{proof}

\begin{theorem}[\complete{} is Sound]
  \label{thm:complete+sound+partial}
  Given a program $s$ that is well typed under typing context
  $\Gamma$, $\Gamma \covervdash s \; : \; b$, and target coverage type
  $\nuut{b}{\psi}$, if \complete{} terminates, it returns a coverage
  complete repaired program $s'$,
  $\complete(\Gamma,~s,~\nuut{b}{\psi}) = s' \implies \Gamma \covervdash s' \; :
  \; \nuut{b}{\psi}$.
\end{theorem}
\begin{proof}
  We proceed by following the structure of the \complete{} algorithm and applying the soundness of its subroutines.

  \textbf{Step 1: Type Inference.} By Lemma~\ref{lemma:typeinfer-sound}, applying $\typeInfer{}(\Gamma, s)$ yields a coverage type $\nuut{b}{\psi_\texttt{cur}}$ such that $\Gamma \covervdash s : \nuut{b}{\psi_\texttt{cur}}$.

  \textbf{Step 2: Abduction.} By \autoref{theorem:generalization+sound}, the call to $\generalize(\Gamma, \nuut{b}{\psi_\texttt{cur}}, \nuut{b}{\psi})$ produces a formula $\psi_\texttt{need}$ such that
  \[
    \Gamma \covervdash \nuut{b}{\psi_\texttt{cur} \lor \psi_\texttt{need}} <: \nuut{b}{\psi}
  \]
  That is, any program covering $\psi_\texttt{cur} \lor \psi_\texttt{need}$ is a subtype of the target coverage $\psi$.

  \textbf{Step 3: Localization.} By Lemma~\ref{lemma:localize-sound}, the call to $\localize(\Gamma, s, \nuut{b}{\psi_\texttt{need}})$ produces a program sketch $s'$ with holes, and for any choice of terms $\overline{e_j}$ such that $\overline{\Gamma \vdash e_j : \nuut{b}{\psi_j}}$, the filled program $s'[\overline{e_j}]$ satisfies
  \[
    \Gamma \vdash s'[\overline{e_j}] : \nuut{b}{\psi_\texttt{cur} \lor \psi_\texttt{need}}
  \]

  \textbf{Step 4: Synthesis.} By Lemma~\ref{lemma:synthesize-sound},
  the call $\synthesize(\Gamma, s', %
  \overline{\Gamma_j\vdash\hole{}_j~:~\nuut{b}{\psi_j}}, %
  \psi_\texttt{cur} \lor
  \psi_\texttt{need})$ will produce a term $s'[\overline{e_j}]$ such
  that $\Gamma \vdash s'[\overline{e_j}]~:~\nuut{b}{\psi_\texttt{cur} \lor
    \psi_\texttt{need}}$.

  \textbf{Step 5: Subsumption.} By the typing subsumption rule and the result of Step 2, since
  \[
    \Gamma \vdash s[\overline{e_j}] ~:~ \nuut{b}{\psi_\texttt{cur} \lor \psi_\texttt{need}}
    \quad \text{and} \quad
    \Gamma \covervdash \nuut{b}{\psi_\texttt{cur} \lor \psi_\texttt{need}} <: \nuut{b}{\psi}
  \]
  we conclude
  \[
    \Gamma \vdash s[\overline{e_j}] ~:~ \nuut{b}{\psi}
  \]
  as required.

  Therefore, if \complete{} terminates, it returns a program $e$ such
  that $\Gamma \vdash e : \nuut{b}{\psi}$, establishing the soundness
  of \complete{}.
\end{proof}

The completeness proofs for \complete{} are based on the following four lemmas.

\begin{lemma}[Termination of $\typeInfer{}$]
  For a given type context $\Gamma$ and incomplete program $s$, $\typeInfer{}$ always terminates.
  \label{lemma:typeinfer-termination}
  \end{lemma}
  \begin{proof}
  This follows from the type completeness theorem for coverage types~\cite{poirot}.
  \end{proof}

\begin{lemma}[Termination of $\generalize{}$]
  For a given type context $\Gamma$, coverage type
  $\nuut{b}{\psi_\texttt{cur}}$ and $\nuut{b}{\psi}$, if there exists
  $\nuut{b}{\psi_\texttt{need}}$ such that
  $\Gamma \covervdash \nuut{b}{\psi_\texttt{cur} \lor
    \psi_\texttt{need}} <: \nuut{b}{\psi}$, $\generalize{}$ always
  terminates.
    \label{lemma:generalize-termination}
    \end{lemma}
\begin{proof}
  The main body of the \generalize{} subroutine is a refinement
  loop. We need to show that this loop terminates. The loop assigns
  each Boolean combination from the set $\Psi^?$ either to $\Psi^+$
  (line 10) or to $\Psi^-$ (line 8). The set $\Psi^?$ is finite
  because it is derived from a finite set of atomic formulas
  ($\Phi$). Thus, the \generalize{} subroutine terminates.
\end{proof}

\begin{lemma}[Termination of $\localize{}$]
  For a given type context $\Gamma$, incomplete program $s$ that
  covers $\nuut{b}{\psi_\texttt{cur}}$, and coverage type
  $\nuut{b}{\psi_\texttt{need}}$, if there exists a program sketch
  $s'$ such that
  $\Gamma \vdash s' : \nuut{b}{\psi_\texttt{cur} \lor
    \psi_\texttt{need}}$, $\localize{}$ always terminates.
    \label{lemma:localize-termination}
    \end{lemma}
    \begin{proof}
      By induction on the structure of the program $s$.
\end{proof}

\begin{lemma}[Termination of $\synthesize{}$]
  For a given type context $\Gamma$ and program sketch $s'$ and terms
  $\overline{e_j}$ where
  $\overline{\Gamma \vdash e_j : \nuut{b}{\psi_j}}$, $\synthesize{}$
  always terminates.
  \label{lemma:synthesize-termination}
  \end{lemma}
  \begin{proof}
    The main body of the \synthesize{} subroutine is an iterative loop
    with an increasing cost bound. Each step of the loop is guaranteed
    to terminate, as there is a finite number of holes,
    $\overline{\hole{}_j}$, and $\genExp$ will only generate a finite
    number of terms at each cost level. The entire loop must
    eventually hit the upper cost bound and terminate. The subsequent
    loop also iterates over at most $j$ unfilled holes, and will also
    terminate. Therefore, \synthesize{} always terminates.
\end{proof}

\begin{theorem}[\complete{} Terminates]
  \label{thm:complete+total}
  \complete{} always terminates.
\end{theorem}
\begin{proof} This is a direct consequence of the previous three
  lemmas:
  \begin{itemize}
\item By Lemma~\ref{lemma:typeinfer-termination}, $\typeInfer{}$ always terminates for any input $s$ and context $\Gamma$.
\item Since a valid repair exists, there must exist a non-bottom type $\nuut{b}{\psi_\texttt{need}}$ such that $\Gamma \covervdash \nuut{b}{\psi_\texttt{cur}} \lor \nuut{b}{\psi_\texttt{need}} <: \nuut{b}{\psi}$. By Lemma~\ref{lemma:generalize-termination}, $\generalize{}$ always terminates.
\item Given $s'$ and $\nuut{b}{\psi_\texttt{need}}$, Lemma~\ref{lemma:localize-termination} ensures that $\localize{}$ always terminates.
\item Given $s'$ and the terms $\overline{e_j}$, Lemma~\ref{lemma:synthesize-termination} ensures that $\synthesize{}$ always terminates.
\end{itemize}
Therefore, \complete{} always terminates.
\end{proof}

\begin{theorem}[\complete{} is Sound and Total]
  \label{thm:complete+sound+total}
  Given a program $s$ that is well typed under typing context
  $\Gamma$, $\Gamma \covervdash s \; : \; b$, and target coverage type
  $\nuut{b}{\psi}$, \complete{} returns a coverage complete generator,
  $\Gamma
  \covervdash \complete(s,~\Gamma,~\nuut{b}{\psi}) \; : \; \nuut{b}{\psi}$.
\end{theorem}
\begin{proof}
  Follows immediately from \autoref{thm:complete+sound+partial} and \autoref{thm:complete+total}.
\end{proof}

\newpage
\section{Unique generated values}
\label{sec:tech+unique}

\begin{filecontents*}{fig/data/unique_data.csv}
Generator,Unique_Count,Total_Duplicates,Duplicate_Count,None_Count
Sized List,16759,3241,1,0
Unique List,18939,1061,1,0
Even List,20000,0,0,0
Sorted List,778,1030,1,18192
Duplicate List,18875,1125,1,0
Red-Black Tree,17921,2079,1,0
Complete Tree,18441,1559,1,0
Sized Tree,15124,4876,1,0
\end{filecontents*}
\csvautotabular[respect all]{fig/data/unique_data.csv}

The table above presents an analysis of the number of unique and duplicate values produced by the Cobb-repaired generators from Figure 11. Each value that is output precisely once is reported as a unique element under the `Unique\_Count' column. The `Total\_Duplicates' column includes the total number of repeated values.  `Duplicate\_Count' reports the number of values that are repeated more than once. Note that for most benchmarks the last value is 1, as the only duplicate value generated tends is always either the leaf node or the empty list: Even List has a non-nil base case and therefore doesn't admit any duplicates. Finally, the `None\_Count' column presents the number of runs on which the generator threw an error -- as described in Section the Sorted List benchmark is the only one that does so.

\end{document}